%% file: paper.tex
\documentclass[]{aa}  
\pdfoutput=1
\usepackage{graphicx}
\usepackage{caption}
\usepackage{subcaption}
\usepackage{epsf}
\usepackage{longtable,lscape}
\usepackage{dcolumn}
\usepackage{booktabs} 
\usepackage{url}
\usepackage{float}
\usepackage{microtype}
\usepackage[varg]{txfonts}
\usepackage{natbib,twoopt} 
\usepackage{array,multirow}
\usepackage{xcolor}
\input{bibdefinitions.tex}

\bibpunct{(}{)}{;}{a}{}{,} % citation like A&A style

\let\oldhat\hat

\renewcommand{\hat}[1]{\oldhat{\mathbf{#1}}}
\newcommand{\kms}{km\,${\rm s}^{-1}$}

\newcommand{\myr}{\mbox{$M_\odot\,{\rm yr}^{-1}$}}

\newcommand{\Mone}{$M^{\rm WR}_{\rm spec}$}
\newcommand{\Mtwo}{$M^{\rm WR}_{\rm single}$}
\newcommand{\Lone}{$L^{\rm WR}_{\rm spec}$}

\newcommand{\HeII}{He\,{\sc ii}\,$\lambda 4686\,$}

\begin{document}

   \title{Why binary interaction does not necessarily dominate the formation of Wolf-Rayet stars at low metallicity}

%    \subtitle{and the forgotten parameter}

   \author{T. Shenar\inst{1} 
          \and A.\ Gilkis\inst{2}     
          \and J.\ S.\ Vink\inst{3}  
          \and H.\ Sana\inst{1}            
          \and A.\ A.\ C.\ Sander\inst{3}             
%           \and R.\ Hainich\inst{3}
%           \and H.\ Todt\inst{2}                            
%           \and A.\ F.\ J.\ Moffat\inst{4}   
%           \and L.\ M.\ Oskinova\inst{2}                 
%           \and V.\ Ramachandran\inst{2} 
%           \and H.\ Sana\inst{1}                     
%           \and A.A.C Sander\inst{5}       
%           \and O.\ Schnurr\inst{6}    
%           \and N.\ St-Louis\inst{4}             
%           \and D.\ Vanbeveren\inst{7}          
%           \and W.-R.\ Hamann\inst{2}           
%           \and W.-R.\ Hamann\inst{1}            
%           \and J.\ J.\ Eldridge\inst{3}
%           \and H.\ Pablo\inst{2}   
%           \and N.\ D.\ Richardson\inst{4}         
          }

   \institute{\inst{1}{Institute of Astrophysics, KU Leuven, Celestijnlaan 200D, 3001 Leuven, Belgium}\\  
              \inst{2}{Institute of Astronomy, University of Cambridge, Madingley Road, Cambridge, CB3 0HA, UK} \\
              \inst{3}{Armagh Observatory, College Hill, BT61 9DG, Armagh, Northern Ireland}\\
              \email{tomer.shenar@kuleuven.be}   
              }
   \date{Received ? / Accepted ?}

%-------------------  Abstract --------------------

\abstract 
{
Classical Wolf-Rayet (WR) stars are massive, hydrogen depleted, post main-sequence stars that exhibit emission-line 
dominated spectra. For a given metallicity $Z$, 
stars exceeding a certain initial mass \Mtwo(Z)~can reach the WR phase through intrinsic mass-loss or eruptions (single-star channel).
In principle, stars of lower masses can reach the WR phase via stripping through binary interactions (binary channel).
Because winds become weaker at low $Z$, it is commonly assumed that the binary channel dominates the formation of WR stars in 
environments with low metallicity such as the Small and Large Magellanic Clouds (\object{SMC}, \object{LMC}).  
However, the reported WR binary fractions of $30-40\%$ in the SMC ($Z=0.002$) and LMC ($Z=0.006$) are comparable to that of the 
Galaxy ($Z=0.014$), and
no evidence for the dominance of the binary channel at low $Z$ could be identified observationally. 
Here, we explain this apparent contradiction  by 
considering the minimum initial mass \Mone(Z)~needed for the stripped product to appear as a WR star. 
}
{By constraining \Mone(Z)~and \Mtwo(Z), we estimate the importance of binaries in forming WR stars as a function of $Z$. 
}
{We calibrate \Mone(Z) using the lowest-luminosity WR stars in the Magellanic Clouds and the Galaxy. A range of \Mtwo~values are explored using 
various evolution codes.  We estimate the additional contribution of the binary channel  by 
considering the  interval [\Mone(Z), \Mtwo(Z)], which characterizes the initial-mass range in which the binary channel can form 
additional WR stars. }
{The WR-phenomenon ceases below luminosities of $\log L{\approx}4.9, 5.25,$ and $5.6\,[L_\odot]$ in the Galaxy, 
the LMC, and the SMC, respectively, which translates to minimum He-star masses of $7.5, 11, 17\,M_\odot$  
and minimum initial masses of \Mone = $18, 23, 37\,M_\odot$. Stripped stars with lower initial masses in the respective 
galaxies would tend to not appear as WR stars.
The minimum mass necessary for self-stripping, \Mtwo(Z), is strongly model dependent, 
but lies in the range $20-30$, $30-60$, and $\gtrsim 40\,M_\odot$ for the Galaxy, LMC, and SMC, respectively. 
% Binary stripping can only form additional WR stars in the mass interval [\Mone, \Mtwo].
We find that
that the additional contribution of the binary channel is a non-trivial and model-dependent function of $Z$ that 
cannot be conclusively claimed to be monotonically increasing with decreasing $Z$.
}
{
The WR spectral appearance arises from the presence of strong winds. Therefore, both \Mone~and \Mtwo increase with 
decreasing metallicity. Considering this, we show that one should not a-priori expect that 
binary interactions become increasingly important in forming WR stars at low $Z$, or that the WR binary fraction 
grows with decreasing $Z$.
}

\maketitle

\section{Introduction}
\label{sec:introduction}

% Massive stars ($M_{\rm i} \gtrsim 8\,M_\odot$) 

The existence of massive stars ($M_{\rm i} \gtrsim 8\,M_\odot$) that have been stripped 
off their outer, hydrogen-rich layers, is implied through  observations of 
stripped core-collapse supernovae \citep[type Ibc SNe, e.g.,][]{Smartt2009, Prentice2019}. However, hydrogen-depleted massive stars are also directly observed. 
Probably the best known and most commonly observed type of stripped massive stars are the classical Wolf-Rayet (WR) stars.

Massive WR stars comprise a spectral class of hot ($T_* \gtrsim 40\,$kK) massive stars with powerful stellar winds that  
give rise to emission-line dominated spectra \citep[for a review, see][]{Crowther2007}.
These radiatively driven winds induce 
typical mass-loss rates in the range of $ -6.0  \lesssim \log \dot{M} \lesssim -3.5\,$[\myr], reaching terminal velocities $v_\infty$ 
of the order of 2\,000\,\kms.
% \approx 1000-2000\,$\kms.
WR stars come in three flavors that correspond to the amount of stripping the star experienced, starting with
nitrogen-rich WR stars (WN), followed by carbon-rich WR stars (WC), and ending with the very rare oxygen-rich 
WR stars (WO). Evolutionarily, one distinguishes between classical WR (cWR) stars, which have evolved 
off the main sequence, and main-sequence WR stars (often classified as WNh, and dubbed ``O stars on steroids''), 
which are very massive stars that possess strong mass-loss already on the main sequence \citep{deKoter1997}. About 90\% of the known WR stars are  cWR 
stars \cite[see Sect.\,1 in ][]{Shenar2019}. cWR stars probe some of the least understood phases 
of massive stars prior to their core-collapse into neutron stars or black holes \citep[e.g.,][]{Langer2012}. 
In this paper, we focus on the relative role of binary interactions
in forming cWR stars.
% in different metallicity environments.

Envelope stripping is an essential ingredient in the formation of cWR stars. 
The loss of envelope increases the   
luminosity-to-mass ratio ($L/M$). The resulting proximity to the Eddington limit enables the launching of 
a powerful wind \citep[e.g.,][]{CAK1975, Graefener2011}. The process escalates 
when the wind becomes optically thick and enters the multiple scattering regime, 
where photons are rescattered multiple times before escaping \citep[e.g.,][]{Lucy1993, Vink2011eta}. The winds of 
WR stars are often found in the multiple scattering regime \citep[e.g.][]{Hamann1993, Vink2012}.

\begin{figure}[!htb]
\centering
  \includegraphics[width=0.5\textwidth]{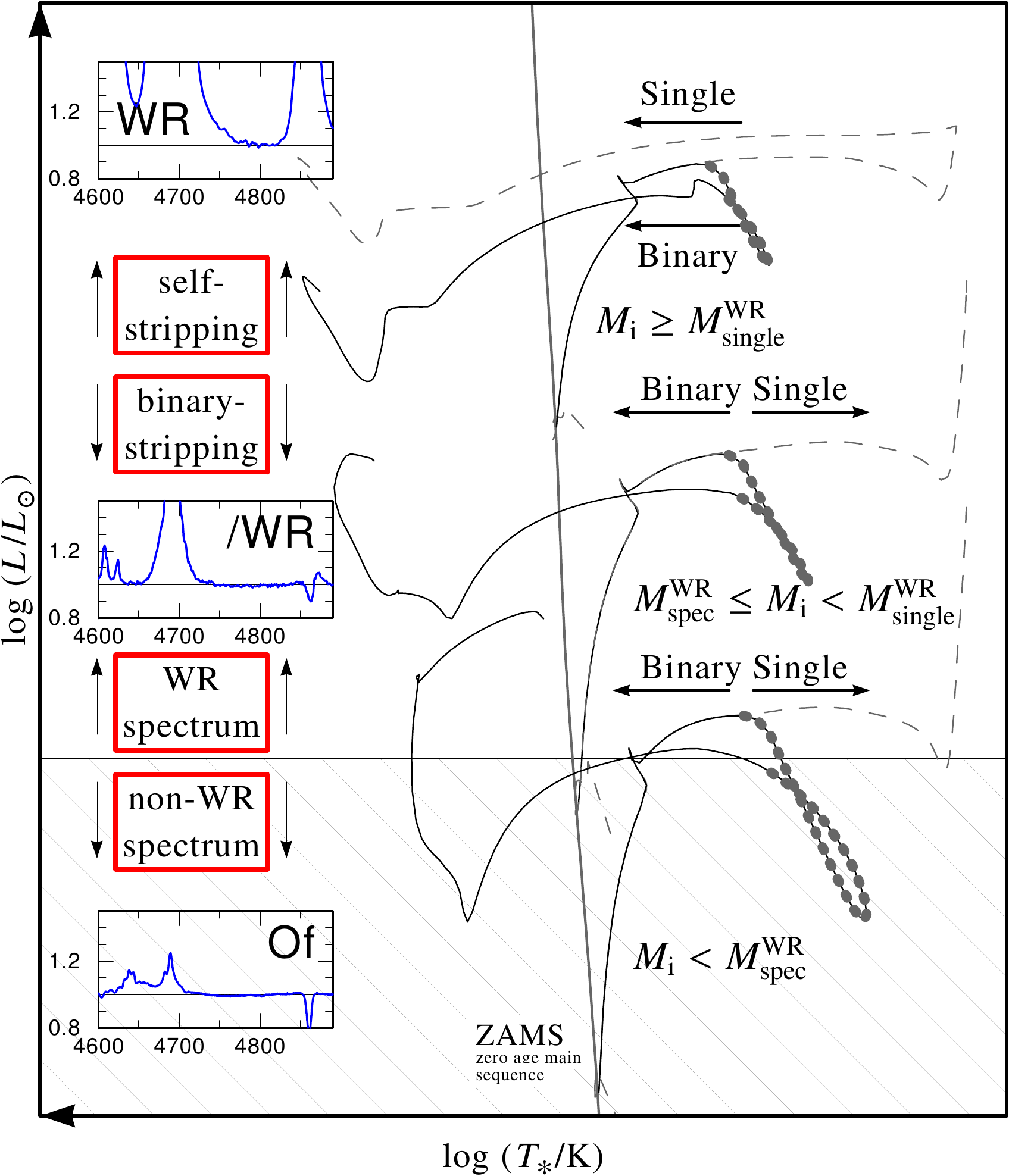}
  \caption{Illustration of the minimum initial mass required to appear as a WR star after stripping (\Mone) and to reach 
  the WR phase through intrinsic stripping (\Mtwo). 
  Plotted are six BPASS evolution tracks at a fixed metallicity ($0.008$), three for single stars and three for binaries with 
  an initial mass ratio of 0.9 and an initial period of $P_{\rm i} =100\,$d. The initial mass of the single star 
  or the primary is fixed to three 
  distinct values, $M_{\rm i} <$ \Mone, \Mone $< M_{\rm i} <$\Mtwo, and $M_{\rm i} > $\Mtwo. The dotted segments correspond 
  to binary mass transfer phases. Only stars with $M_{\rm i} > $\Mtwo~can undergo 
  self-stripping. Stars with  $M_{\rm i} <$\Mtwo~can only be stripped in binaries, but only those with 
  $M_{\rm i} $>\Mone~will appear as WR stars after stripping.  
  The insets represent possible normalized spectra of He stars found in the corresponding parameter regime, 
  noting that significant deviations for individual objects are possible.
  Since the purpose of the figure is illustrative, the 
  values of the initial masses and axis labels (which are model and $Z$-dependent) are omitted. 
  }
\label{fig:M1M2illustration}
\end{figure} 

It was originally thought that mass-transfer in binaries is responsible for the removal 
of the H-rich envelope and the formation of 
cWR stars \citep{Paczynski1967}. It was only later, 
with the realization that massive stars can drive significant winds,
that stripping through winds 
or eruptions associated with single stars became a viable
alternative \citep{Conti1976, Abbott1987, Smith2014}.
The more massive a star is, the stronger its mass-loss rate and the larger its convective core.
Thus, it is expected that stars with initial masses exceeding a certain threshold (at a given metallicity $Z$) will 
be able to undergo self-stripping. 
We denote this threshold mass as \Mtwo~(Fig.\,\ref{fig:M1M2illustration}): the minimum mass necessary for a single 
star to reach the cWR phase.

Once a theoretical framework of stellar winds was established, 
the single-star WR channel  (dubbed the Conti scenario after \citealt{Conti1976}) took precedence over 
the binary channel. However, since then, it became clear that the majority of massive stars interact with a companion during their 
lifetime \citep{Sana2012, Sota2014}.
Moreover, the inclusion of clumping formalisms in atmosphere models resulted in a systematic lowering of $\dot{M}$ 
values for massive O-type stars \citep{Puls2006, Fullerton2006}. 
This gave rise to a renewed discussion regarding the importance of binary mass-transfer to the formation 
of WR stars.  

Owing to the dependence of the opacity on the metallicity, 
mass-loss rates of massive stars decrease with $Z$ \citep{Vink2001, Vink2005, Crowther2006, Hainich2015, Shenar2019}. 
The immediate conclusion is that, the lower the metallicity, the harder it would be for a star to 
undergo self-stripping and become a   
cWR star.  Hence, the efficiency of the single-star channel drops with $Z$.
In contrast, the efficiency of binary stripping appears to 
be largely $Z$-independent: no evidence exists that the binary frequency of massive stars strongly depends on $Z$
\citep{Sana2013, Dunstall2015}, and if any, 
it rather points towards an increased fraction of close binaries for low-mass stars at low $Z$ \citep[e.g.,][]{Moe2019}. 
% While the mass-transfer efficiency  
% may depend on the metallicity, 
It would appear naively that the  fraction of WR stars forming through binary mass-transfer should grow with decreasing metallicity. 
Indeed, 
this expectation is repeatedly claimed in the  literature \citep[e.g.,][]{Maeder1994, Bartzakos2001, Smith2014, Groh2019}.

The Small and Large Magellanic Clouds (SMC, LMC), with metallicity
contents of $Z\approx 0.2, 0.4\,Z_\odot$ \citep[e.g.][]{Hunter2007} 
serve as precious laboratories to study the formation of WR stars at low $Z$.  
Surveys conducted by \citet{Massey2003, Massey2014} and \citet{Neugent2018} have revealed 
a total of 12 WR stars in the SMC and 154 WR stars in the LMC - samples that are considered to be largely complete. 
\citet{Bartzakos2001}, \citet{Foellmi2003SMC, Foellmi2003LMC}, and \citet{Schnurr2008}  attempted to measure the 
fraction of close ($P \lesssim 100\,$d) WR binaries in the SMC and LMC, 
and reported binary fractions of $\approx$40\% and 30\%, respectively.
These fractions are broadly consistent with the reported \object{Milky Way} (MW) WR binary fraction of 40\% \citep{Vanderhucht2001}, as 
well as with those reported for the super-solar environments of the galaxies M31 and M33
\citep{Neugent2014}. 
\citet{Hainich2014, Hainich2015} and \citet{Shenar2016, Shenar2017, Shenar2018, Shenar2019} performed 
a spectral analysis of the apparently-single and binary (or multiple) WN stars 
in the SMC and the LMC. Contrary to expectation, the distribution of the apparently-single and binary WN stars on the Hertzsprung-Russell 
diagram (HRD) was found to be comparable. 
% Similar findings were reported for a sample of Galactic 
% binary and single WR stars \citep[e.g.,][]{Massey1981}. 
No evidence could be found that the importance of the binary channel grows with decreasing $Z$.

%https://arxiv.org/abs/1602.06358 (MENTION!!! EXAMPLE FOR OVERLOOKING WR DEFINITION)

In this work, we aim to explain why this may be the case. We show that the prediction that binary interactions become increasingly important 
in forming WR stars at low $Z$ lies on a fallacy that is rarely considered in this context. For a star to exhibit a WR spectrum, it needs 
to have a significant stellar wind. Thus, while stars of arbitrary masses can be stripped through binaries and become He stars, 
not all He stars would appear as WR stars. A certain $Z$-dependent threshold exists for the initial 
mass below which the stripped He star would not appear as a WR star. We mark this threshold as \Mone.
In Fig.\,\ref{fig:M1M2illustration}, we illustrate the meaning of \Mone~and \Mtwo~using evolution tracks calculated for single and binary 
stars at a fixed metallicity of three different masses using the BPASS\footnote{bpass.auckland.ac.nz} (Binary Population and Spectral Synthesis) code  V2.0  
\citep{Eldridge2008, Eldridge2016}. 

We argue that both \Mone~and \Mtwo, which will be defined more precisely in Sect.\,\ref{sec:Defs}, increase with decreasing $Z$, 
and that binary interactions can primarily affect the number of WR stars in the initial mass interval \mbox{$M_{\rm i} \in$ [\Mone, \Mtwo]}. Stars outside this interval 
will either not look like WR stars after stripping, or would be massive enough to undergo self-stripping.
By estimating \Mone~(Sect.\,\ref{sec:EstimatingM1}) and \Mtwo~(Sect.\,\ref{sec:EstimatingM2}) for different metallicities, we show that 
the prediction that the binary formation channel should become increasingly important in forming WR stars at low $Z$ 
is not supported by our results, providing an explanation for the apparent independence of the WR binary fraction on $Z$
(Sect.\,\ref{sec:bincont}).
% We conclude by providing appropriate criteria for the classification 
% of stripped stars as WR stars in evolution models, which are calibrated against empirical populations of WR stars (Sect.\,\ref{sec:discussionB}).

\section{Definitions}
\label{sec:Defs}

\subsection{WR stars, He stars, and binary-stripped stars}

\begin{figure}[!htb]
  \includegraphics[width=\linewidth]{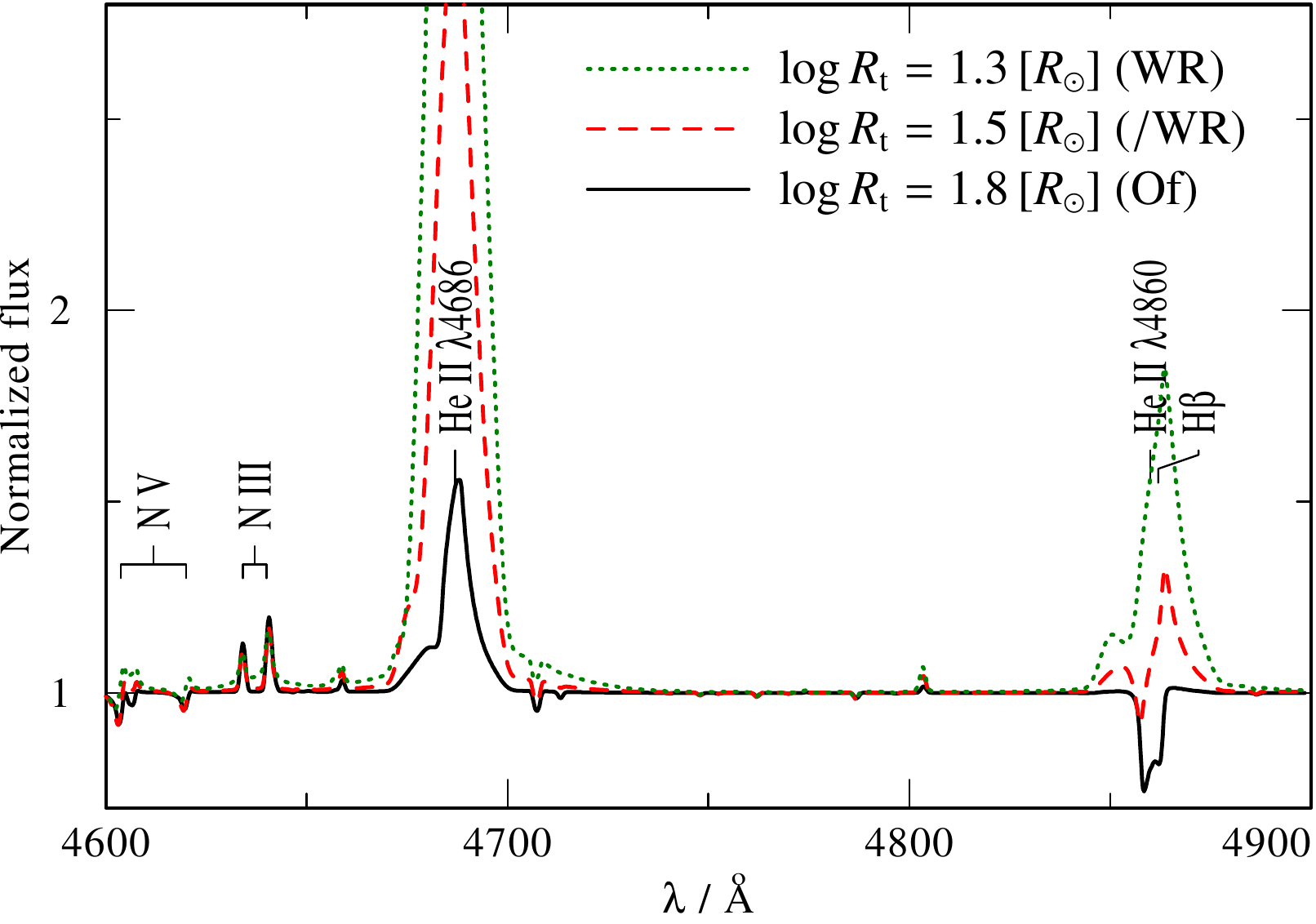}
\caption{Synthetic PoWR spectra calculated for fixed parameters 
($T_*{=}50\,$kK, $\log L{=}5.3\,[L_\odot]$, $M{=}13\,M_\odot$, 
$v_\infty{=}1000\,$\kms, $D{=}10$, surface hydrogen mass fraction $X_{\rm H}{=}0.4$, 
LMC-like composition) but different $R_{\rm t}$ (see legend). 
The figure illustrates that stars with $R_{\rm t}$ values above 
a certain threshold would not appear as WR stars.
}
\label{fig:Rtmods}
\end{figure}

There is a great deal of confusion between WR stars and so-called ``helium stars'' and ``stripped stars''.
However, WR stars comprise a purely spectroscopic class of stars,  much like O or G stars. Loosely speaking, 
WR stars are stars that show broad emission lines of
typical widths of a few hundreds to thousands of \kms
\citep[see, e.g.,][for an early characterization]{Beals1940}. Stars having such a spectral appearance are often 
associated with cWR stars, but they can also be very massive main sequence stars, or central stars of planetary
nebula ([WR] stars).

The distinction between binary-stripped stars, He stars, and WR stars is important to respect. Otherwise, 
comparing observed populations of WR stars with predictions can become confusing and meaningless. 
To make sure we are ``comparing apples with apples'', 
it is therefore best to abide by the observer's definition of WR stars: \emph{A star is a WR star if it would be recognized as such in 
WR surveys}. Relying on common nomenclature, we define: 
\vspace{-.2cm}
\begin{itemize}
 \item \emph{He stars:} core He-burning, H-depleted stars 
 \item \emph{binary-stripped stars:} He stars that were stripped in binaries
 \item \emph{classical WR stars:} He stars with a WR spectrum, whether stripped through intrinsic mass-loss or binaries.
\end{itemize}
Thus, binary-stripped stars and  cWR stars are always He stars, but the converse does not hold. 

The discovery of WR stars relies on image subtractions taken with narrow-band filters \citep[see recent review by][]{Neugent2019}. The filters are designed to focus on the 
most prominent 
emission features of WR stars. For WN stars, the filters focus on the \HeII line, and since 
% , while for WC stars, they focus around the \CIIIfilter line. Since
WN stars are generally the 
first and longest-lived phase of WR stars, we concentrate here on WN stars. To be recognized as a WN star, 
the \HeII emission line should typically reach at least twice the continuum 
level in peak intensity. After being identified through narrow-band filters, 
spectra are usually taken to confirm a WR-like spectrum. The current 
convention is that the spectrum needs to show some signs of emission in the H$\beta$ spectral range (which includes the blended 
He\,{\sc ii}\,$\lambda 4860$ line) for the 
star to be considered a WR star. Stars in which the H$\beta$ line is only partially filled with emission (red dashed line in Fig.\,\ref{fig:Rtmods})
are known as ``slash'' WR stars, 
which are  classified either as O/WR or WR/O in the literature \citep{Crowther2011, Neugent2017}. 
Other hot stars that do not show emission in H$\beta$ but some emission in \HeII are  
typically classified as Of stars (see Fig.\,\ref{fig:Rtmods}).

To exhibit significant wind emission, a star needs to possess a strong wind. But what is ``strong''? 
Experience shows that the WR stars\footnote{Throughout this paper, only massive stars and their WR counterparts are considered. The 
low-mass central WR stars of planetary nebula - so called [WR] stars - obey a different mass-luminosity relation
\citep[e.g.,][]{Leuenhagen1998}, and are not regarded here.}
with the weakest emission lines have mass-loss rates 
of $\log \dot{M} \approx -6.0\,$[\myr]. 
It is important to remember, however, that the strength of the emission also depends on the 
size of the stellar surface, which determines the continuum level.
A helpful parameter that quantifies 
the relative strength of the emission lines for a given temperature and surface composition
is the so-called transformed radius $R_{\rm t}$ \citep{Schmutz1989}:
\begin{equation}
 R_\text{t} = R_* \left[ \frac{v_\infty}{2500\,{\rm km}\,{\rm s}^{-1}\,}  \middle/  
 \frac{\dot{M} \sqrt{D}}{10^{-4}\,M_\odot\,{\rm yr}^{-1}}  \right]^{2/3},
\label{eq:Rt}
\end{equation}
where $R_*$ is the stellar radius, $v_\infty$ is the terminal velocity, and
$D$ the clumping factor. The way it is defined, smaller $R_{\rm t}$ values correspond to stronger emission lines. 
At a given temperature and surface composition, stars with $R_\text{t}$ values that exceed  
a certain threshold will not appear as WR stars. This is illustrated in Fig.\,\ref{fig:Rtmods}, where 
we show three models calculated with the Potsdam Wolf-Rayet (PoWR) model atmosphere code
\citep{Graefener2002, Hamann2003, Sander2015} for 
He stars with fixed parameters (see caption of Fig.\,\ref{fig:Rtmods}) but different $R_{\rm t}$ values, corresponding 
to WR, /WR, and Of spectral appearances.

% We show that carefully considering the definition of WR stars leads to surprising results regarding previous predictions about the 
% role of binaries in forming WR stars.

\begin{figure*}[!htb]
\minipage{0.32\textwidth}
  \includegraphics[width=\linewidth]{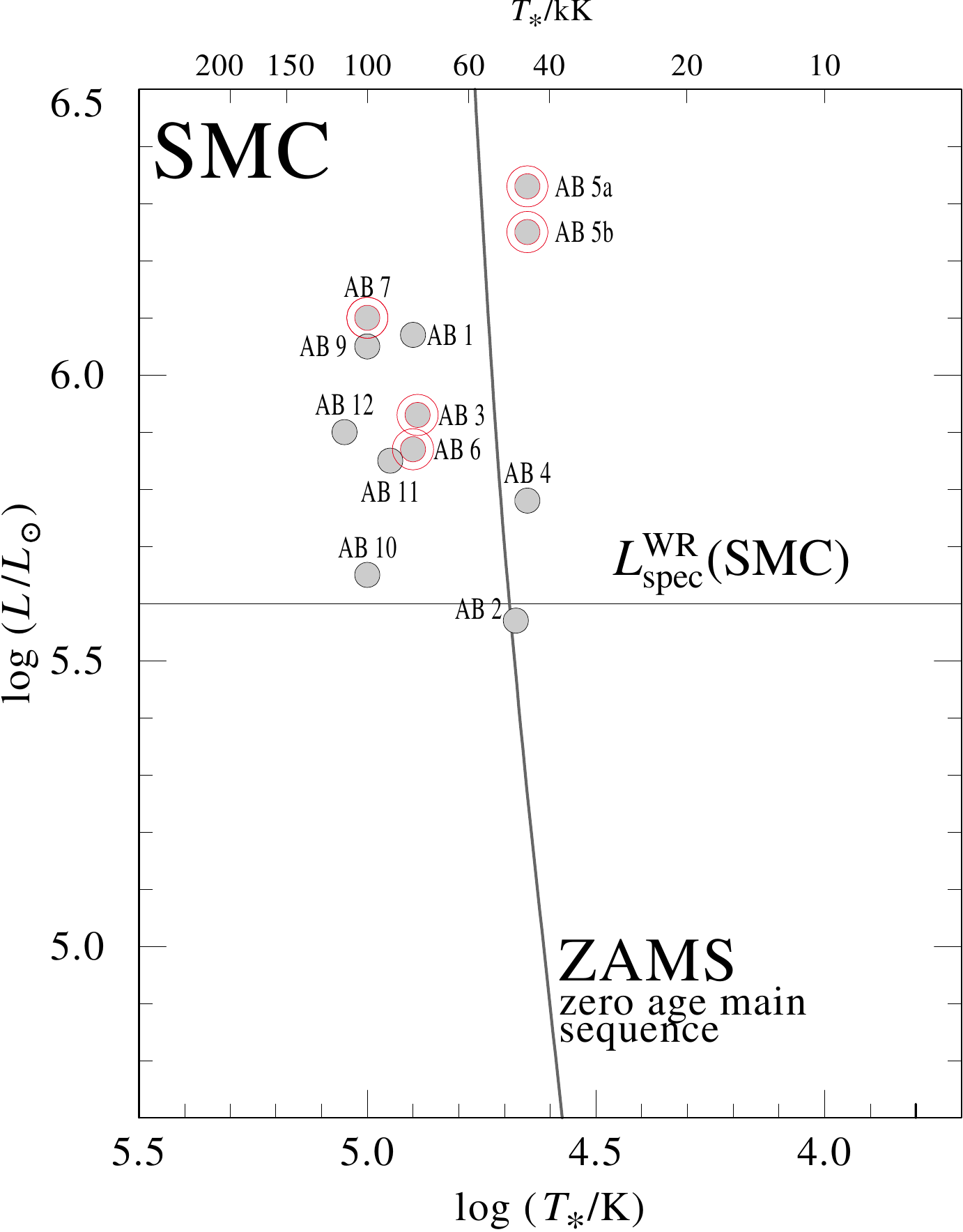}
%   \caption{A really Awesome Image}
%   \label{fig:awesome_image1}
\endminipage\hfill
\minipage{0.32\textwidth}
  \includegraphics[width=\linewidth]{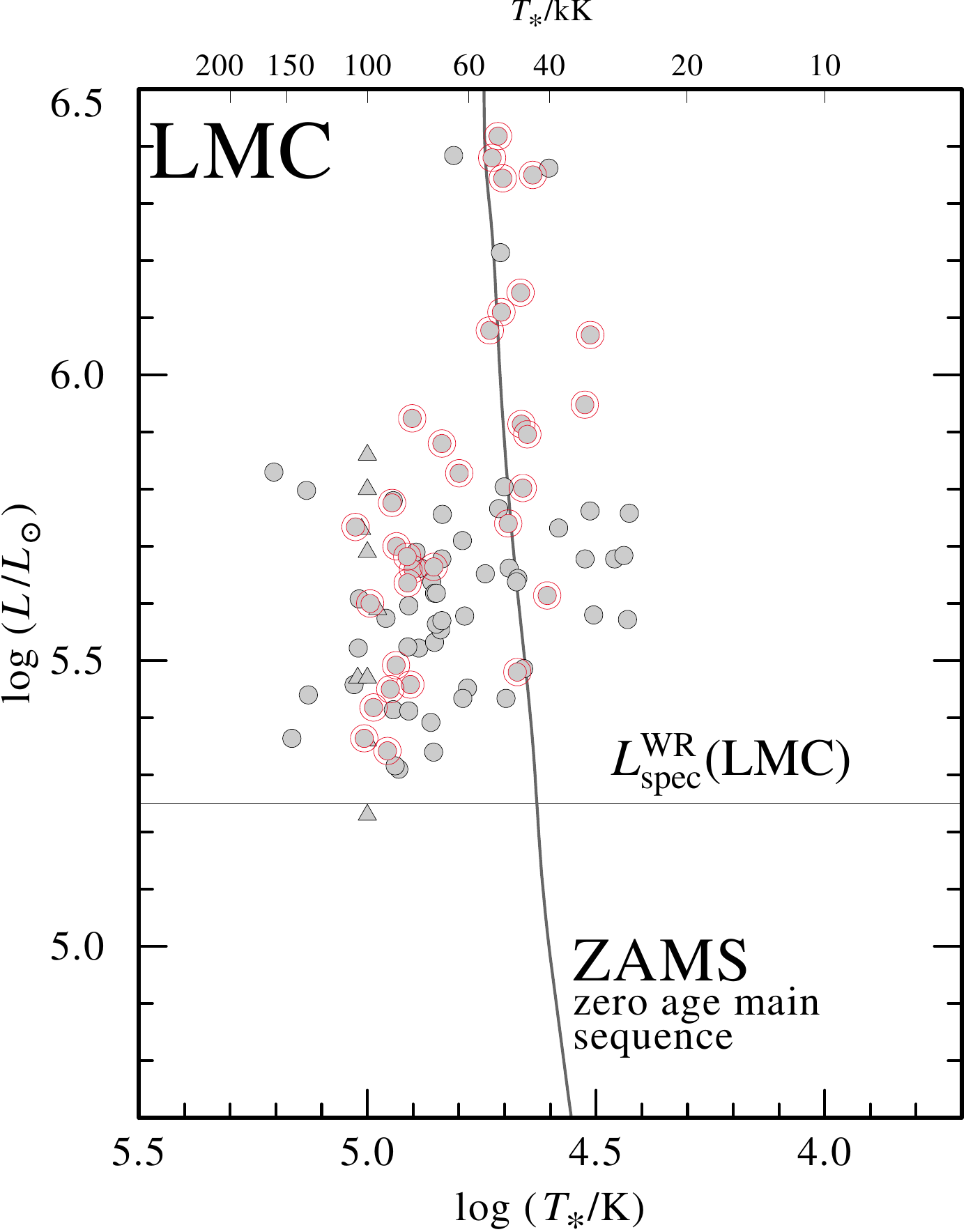}
%   \caption{A really Awesome Image}
%   \label{fig:awesome_image2}
\endminipage\hfill
\minipage{0.32\textwidth}%
  \includegraphics[width=\linewidth]{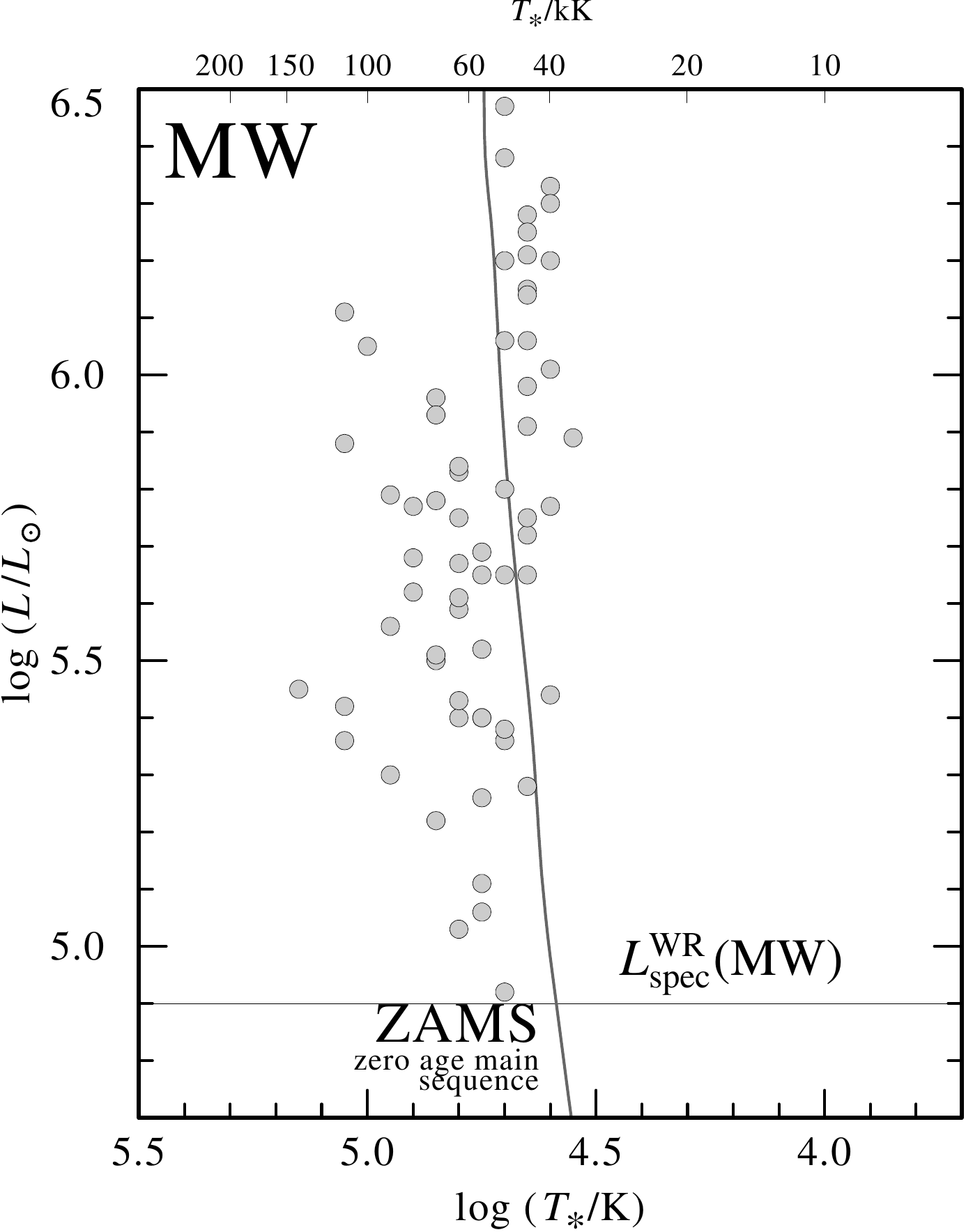}
%   \caption{A really Awesome Image}
%   \label{fig:awesome_image3}
\endminipage
\caption{Observed populations of WR stars in the SMC (left panel), LMC (middle panel), and MW (right panel). The SMC and LMC populations 
include both apparently single WN stars \citep[][gray circles]{Hainich2014, Hainich2015},  
WN binaries \citep[][gray circles surrounded by red circles]{Gvaramadze2014, Shenar2016, Shenar2017, Shenar2018, Shenar2019},
and the so-called LMC WN3/O3 stars \citep[][gray triangles]{Neugent2017}.
The MW sample includes only apparently-single 
WN stars \citep{Hamann2019}.
Marked are the lowest luminosities measured for WR stars in each respective galaxy, rounded to 0.05\,dex,
where a clear metallicity trend is apparent. 
% BPASS binary tracks calculated for the stripped primary are shown for a convenient translation of $L_{\rm min}$ to \Mone.
}
\label{fig:WRpops}
\end{figure*}

\subsection{\Mone~and \Mtwo}

It is common in the literature to consider the minimum mass at which a star would reach the cWR phase as a single star.
However, as outlined in the introduction, in order to correctly estimate the impact of binary interaction in forming WR stars, 
this parameter is not sufficient. Additionally, one must consider 
the minimum mass at which a He star would \emph{spectroscopically appear} as a WR star. 
As we will argue in Sect.\,\ref{sec:EstimatingM1}, below a certain mass threshold, which is $Z$-dependent, 
the stripped product would not appear as a WR star. We therefore define:

\begin{enumerate}
 \item \Mone: The minimum initial mass necessary to appear as a WR star after stripping.
 \item \Mtwo: The minimum initial mass necessary to reach the cWR phase as a single star. 
\end{enumerate}

\Mone~and \Mtwo~are two very different parameters. \Mone~refers to the spectroscopic appearance of the He star, 
which primarily depends on the \emph{current} mass-loss rate of the He star. It does not (directly)  
depend on the evolution history of its progenitor.  In contrast, \Mtwo~is directly 
related to the mass-loss and evolutionary history \emph{prior} to the formation of the cWR star. 
% In the literature, it is \Mtwo~that is typically considered. 
% However, we must not forget that in order to appear as a WR star, the progenitor must fulfil $M_{\rm i} >$\Mone.  
% We show that a careful consideration of this fact results in a significant reduction of the impact of binary interaction in forming WR stars. 

We note that \Mone $\le $\Mtwo~always holds simply from the way \Mone~and \Mtwo~are defined. Per 
definition, if the initial mass \mbox{$M_{\rm i} >$ \Mtwo}, then the star is massive enough to reach the cWR phase as a single star. 
This must mean that the stripped product exhibits a WR spectrum - otherwise one could not claim that it has reached the cWR 
phase. Hence, it must hold that \mbox{$M_{\rm i} >$ \Mone}.
This subtle and rigorous definition ensures that \mbox{\Mone $\le$ \Mtwo}, irrespective of model uncertainties.

\section{Estimating \Mone: What are the initial masses of He stars that appear as WR stars?}
\label{sec:EstimatingM1}

We begin by estimating the minimum initial mass necessary to appear as a WR star after stripping, \Mone. 
In principle, one could calculate model atmospheres and 
determine the mass at which a WR appearance is obtained \citep[e.g.][]{Goetberg2018}. However,
the most important parameter in this context - $\dot{M}$ - is also 
the biggest unknown \citep[e.g.,][]{Gilkis2019}. While $\dot{M}$ is fairly well constrained for WR stars 
\citep{Nugis2000, Hainich2014, Shenar2019}, it is poorly constrained for He stars in the intermediate mass 
range $2-8\,M_\odot$, 
which were hardly ever - perhaps never - directly observed.
\citet{Vink2017} provided prescriptions for such stars, but the 
one peculiar candidate for an intermediate-mass He star - the so-called qWR star \object{HD\,45166} - does not seem to agree with these predictions, 
potentially as this star exhibits a two-component stellar wind \citep{Steiner2005, Groh2008}.
Alternatively, one could rely on hydrodynamically-consistent models for the calculation of $\dot{M}$ \citep{Graefener2005, Sander2017, Sundqvist2019}. However, such models are extremely computationally extensive and suffer from other uncertainties, 
such as the treatment of clumping \citep{Oskinova2007, Sundqvist2018}.  

A more reliable method can instead utilize 
observed WN populations. In Fig.\,\ref{fig:WRpops}, we show the HRD positions of 
the WR populations in the 
SMC, LMC, and MW, adopted from \citet{Hainich2014, Hainich2015}, \citet{Shenar2016, Shenar2017, Shenar2018, Shenar2019}, 
\citet{Neugent2017}, \citet{Gvaramadze2014}, and \citet{Hamann2019}.
The SMC and LMC populations include both apparently-single and binary (or multiple) WN stars, 
and are thought to be largely complete \citep[e.g.,][]{Neugent2018}. 
The MW sample includes only apparently-single WN stars \citep{Hamann2019}, since the binaries 
were not yet systematically analysed. For this reason, and due to the extreme extinction towards
the Galactic center, the MW sample is far from complete. 
We note that the HRD distributions of single and binary WR stars are not  
suggestive of a pure ``binary'' interval [\Mone, \Mtwo], as already pointed 
out by \citet{Shenar2019}. This fact is discussed in Sect.\,\ref{subsec:lackobs}.

Two facts become apparent from these distributions. 
First, that the luminosities of the WN stars do not fall below a certain luminosity threshold  
of at least $\log$ \Lone~$\gtrsim 4.9\,[L_\odot]$.
Second, that this threshold is $Z$-dependent, increasing with decreasing $Z$. Below, we motivate these facts analytically.

\subsection{The existence of the minimum mass \Mone}
\label{subsec:Moneex}

One may think that the lower luminosity limit of WR stars, \Lone, exists because  only sufficiently massive 
stars can strip themselves and become WR stars ($M_{\rm i} >$\Mtwo). 
However, binary interactions can strip stars of arbitrary masses. Why then do we not observe a 
continuum of WR stars to arbitrarily low luminosities? 
The reason lies, we believe, in the fact that 
He stars with initial masses lower than a certain threshold - \Mone - would not 
display a WR spectrum. This fact can be 
motivated analytically.

Whether or not a star appears as a WR star depends primarily on the strength of its wind ($\propto \dot{M}$) relative to its 
continuum ($\propto L$). The transformed radius $R_\text{t}$ encompasses this ratio. Thus, for a given effective temperature 
and surface composition,
stars with $R_\text{t}$ 
values above a certain threshold will tend to not appear as WR stars (Fig.\,\ref{fig:Rtmods}). 
For simplicity, let us assume that 
He stars reach similar effective temperatures and abundance patterns. 
One therefore needs to derive the dependence of $R_\text{t}$ 
on $L$. Assuming the dependence of $v_\infty$ and $D$ on $L$ is negligible, Eq.\,(\ref{eq:Rt}) yields 
\begin{equation}
 R_\text{t} \propto L^{1/2} \cdot \dot{M}^{-2/3}, 
\label{eq:Rtprop}
\end{equation}
where the Stefan-Boltzmann 
relation $R_* \propto L^{1/2}$ was used. The challenge is to express $\dot{M}$ in terms of $L$. 

Several empirical determinations of $\dot{M}$ and $L$ find 
an almost linear relation 
between the two in the WR regime \citep[e.g.][]{Hainich2015, Shenar2019}. 
Theoretical and empirical studies of hydrogen-rich WR stars \citep{Vink2012, Bestenlehner2014}
show that below a certain L, the $\dot{M}-L$ relations reveal a ``kink'' due to the transition into an
optically thin wind regime. 

\begin{figure}[!htb]
\centering
  \includegraphics[width=0.5\textwidth]{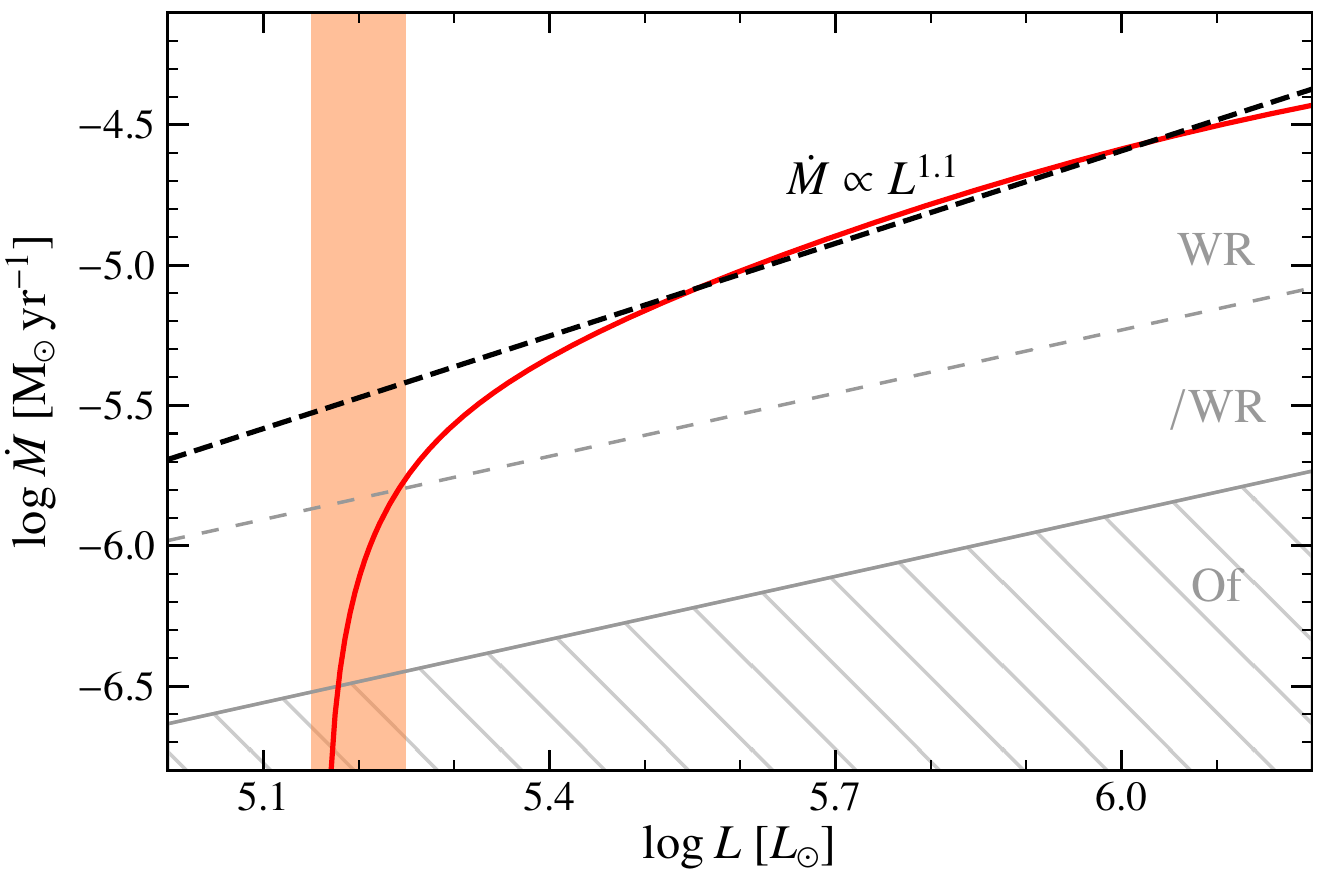}
  \caption{Schematic behaviour of $\dot{M}$ as a function of L for a hydrogen-free WN
star with $T_* = 141$\,kK, based on the series of
hydrodynamcially-consistent WN models at $Z_\odot$ from Sander et al.
(2019) and calibrations with the M(L)-relation from \citet{Graefener2011}. The upper dashed line indicates 
a linear trend in the WR regime, as confirmed from empirical measurements (see text). Lines of 
$R_{\rm t} = {\rm const}$ ($L \propto \dot{M}^{3/4}$, see Eq.\,\ref{eq:Rtprop}) roughly indicate 
the regions in which a WR, /WR, and Of spectral appearance
is obtained. The shaded orange region 
roughly marks the transition to a WR spectral appearance. We note that the curve 
and the position of the lines of $R_{\rm t} = {\rm const}$ 
depend on the choice of $T_*$ and $X_{\rm H}$, and that the purpose of this figure is primarily illustrative.
}
\label{fig:Sander}
\end{figure}

Recent calculations by \citet{Sander2019} reveal that there is
also a transition in the wind regime for hydrogen-free WR stars.
In Fig.\,\ref{fig:Sander}, we show the behaviour of $\dot{M}(L)$ inferred from a series of
hydrodynamically-consistent WN models with $T_*{=}141\,$kK. For this,
we calibrated their $\dot{M}(L/M)$-relation for $Z_\odot$ with the
$L(M)$-relation for He stars from \citet{Graefener2011} in order to
eliminate the explicit mass dependence. A power-law fit to the higher
luminosity range confirms the empirical finding of a near-linear trend
in this regime. However, the models also show that at lower luminosities
(here $\approx 5.2$\,dex) there is a dramatic drop in the mass-loss rate
that coincides with the eventual disappearance of the WR features in
their spectrum. 
The luminosity cut-off in Fig.\ref{fig:Sander} is in agreement with observations of
hot WR stars (see Fig. 3, right panel). However, given the various approximations made to
derive this $\dot{M}(L)$-relation, we do not expect it to exactly yield
the observed threshold, which is set by cooler cWR stars with thin hydrogen mantels. 
To obtain a more accurate estimate, additional studies with
hydrodynamical models accounting for the different temperatures and  surface 
abundances would be required.

In the WR regime, by plugging $\dot{M} \propto L$ in Eq.\,\ref{eq:Rt}, 
one obtains $R_\text{t, WR} \propto L^{-1/6}$. 
The small exponent  implies 
that the dependence of secondary parameters on $L$ (e.g., $v_\infty$) cannot be neglected in the WR regime, and 
hence one should not over-interpret the meaning of such a relation. 
However, once a critical luminosity has been reached, $\dot{M}$ drops abruptly primarily due to the transition 
to an optically-thin wind (Fig.\,\ref{fig:Sander}), corresponding to an abrupt increase in $R_{\rm t}$
(Eq.\,\ref{eq:Rtprop}). This transition in $\dot{M}$ closely coincides with
the disappearance of the WR features from the spectrum, implying that He stars 
below this luminosity threshold would not appear as WR stars. 
Hence, as observed, the WR phenomenon ceases below a relatively sharp luminosity threshold \Lone. 
This luminosity threshold can be translated to a minimum He-star mass, which can in turn 
be translated into a minimum initial mass, $M_{\rm i} = $\Mone. 

We note that this derivation is valid for fixed temperature,  terminal velocity, clumping factor, and composition.
However, these parameters do generally depend on $L$. For example, 
lower-mass binary-stripped stars tend to be somewhat cooler and contain more hydrogen \citep[e.g.,][]{Paczynski1971, Goetberg2018}. 
Nevertheless, since we merely wish to motivate the existence of \Mone~in this section, this has little consequence  
on our final conclusion.

\subsection{\Mone~increases with decreasing $Z$}
\label{subsec:MoneZ}

The fact that \Mone~increases with decreasing $Z$ should also not come as a surprise. 
The mass-loss rates of WN stars were reported to scale with 
$Z$ roughly as $\dot{M}_{\rm WR} \propto Z^{0.8}$ \citep{Vink2005, Hainich2015, Shenar2019}, 
with a shallower dependency reported for WC stars due to their self enrichment in metals  \citep[e.g.,][]{Tramper2016}. 
Even though the $\dot{M}_{\rm WR}-Z$ relation may be more complex in truth, it is clear that, as $Z$ decreases,  
larger He-core masses are needed to generate a WR spectrum. The threshold mass for the spectroscopic 
appearance of a WR star, \Mone, therefore increases with decreasing $Z$.

The  HRD distributions of WR stars in the SMC, LMC, and MW enable us to determine \Mone~virtually without assumptions. 
This is especially the case for the SMC and LMC, in which the WR populations include all known apparently-single and 
binary WN stars. Moreover, 
WR stars close to the lower threshold of \Lone~tend to exhibit weak emission lines blended with 
absorption, which strongly 
indicates that these WR stars are on the ``verge'' of appearing as Of stars. That is, we can be certain that we are sensitive 
to the lowest-luminosity WR stars. 

From the populations shown in Fig.\,\ref{fig:WRpops}, we can draw threshold luminosities of approx.\ 
$\log$\,\Lone = 5.6, 5.25, and 4.9$\,[L_\odot]$ for the SMC, LMC, and MW, respectively. These values are estimated 
from the minimum WR luminosities in the respective galaxies, rounded to 0.05\,dex.
Clearly, the exact value would depend on the strict definition of WR stars, 
and is affected by uncertainties in $\log L$ (typically 
of the order of 0.1\,dex), sample sizes, and completeness aspects.  All in all, the exact values of \Mone~are
bound to be slightly subjective. However, our main argument, which 
will be outlined in Sect.\,\ref{sec:bincont}, is not affected by such uncertainties. 

We now convert these minimum threshold luminosities to minimum masses of WR stars. For this purpose, we use 
mass-luminosity relations published by \citet{Graefener2011}. 
We obtain $M^{WR}_{\rm spec, He} = $ 7.5, 11, and 17\,$M_\odot$. So, for example, a $10\,M_\odot$ He star would appear as a 
WR star in the MW, but not in the SMC.
These masses  
are consistent with dynamical masses measured in WR binaries 
in the MW \citep[e.g.][]{Vanderhucht2001}, the LMC \citep[e.g.][]{Shenar2019}, and the SMC \citep[e.g.][]{Shenar2016}. 

Translating these core masses back to progenitor masses is somewhat 
model-dependent, because it depends on the age of the He star and on the amount 
of mixing and rotation of the progenitor star. We discuss these 
uncertainties in more length in Appendix\,\ref{sec:M1ini}. By estimating 
the most likely progenitor masses associated with the He-star masses derived above, 
we find \Mone{=}18, 24, and 37\,$M_\odot$ for the MW, LMC, and SMC, respectively. 
The inferred \Mone~values for the different metallicities are compiled in Table\,\ref{tab:M1M2}.

One may notice that the values derived above for \Lone~appear to roughly follow \Lone$\propto Z^{-1}$, 
raising the question of whether 
this can be backed by theory.
Let us again assume for simplicity that $v_\infty$, $D$, and $R_*$ do not depend on $Z$ or $L$, 
and that $\dot{M}$ follows a relation in the form $\dot{M} \propto Z^\alpha\,L^\beta$.
Inserting all of this to 
Eq.\,(\ref{eq:Rtprop}), one finds $R_{\rm t} \propto L^{(3-4\beta)/6}\,Z^{-2\alpha/3}$. For constant $R_{\rm t}$ one obtains:\\
\begin{equation}
 L_{\rm spec}^{\rm WR} \propto Z^{4\alpha / (3 - 4\beta)}.
\end{equation}
Our values for \Lone~are calibrated using the lowest-luminosity WR stars, which, at least 
in the SMC and LMC, appear to have optically-thin winds (i.e., they are on the verge of being Of stars).
Unlike typical WR stars, the mass-loss rates of such stars were shown
to follow within considerable scatter
the $\dot{M}-L$ relation derived by \citet{Vink2017} for He stars with optically-thin winds \citep{Shenar2019}. 
Relying on this relation ($\alpha{=}0.61$ and $\beta{=}1.36$), one obtains (at a rather astonishing accuracy) 
\Lone$\propto Z^{-1}$.
However, considering the assumptions performed here (e.g., neglecting $v_\infty$ and $T_*$) and the non-trivial dependence 
of $\dot{M}$ on $Z$ \citep[e.g.,][]{Sander2019}, one must acknowledge that a theoretical prediction of \Lone(Z) 
needs to be confirmed with consistent models, which is beyond the scope of the current paper.

\renewcommand{\arraystretch}{1.2}
\begin{table}[h]
\small 
\setlength\tabcolsep{1.7pt}
\caption{Values for \Mone~and \Mtwo~as a function of $Z$} 
\label{tab:M1M2}
\begin{center}
\resizebox{0.5\textwidth}{!}{
\begin{tabular}{c|c|ccccc}
\hline \hline
  $Z$  &   \Mone \,$[M_\odot]$       &             \multicolumn{5}{|c}{\Mtwo\,$[M_\odot]$}          \\ 
       &        &    BPASS\tablefootmark{a}  & Gen,rot\tablefootmark{b}  & LC18,rot\tablefootmark{c}       &    LC18,non-rot\tablefootmark{d}   &  D03\tablefootmark{e}    \\
\hline                             %
0.014 &  18 & 28  & 23    &  18\,\tablefootmark{f}   & 23 & 33  \\
% \hline               
0.006 &  24 & 33  & 75   &  28 & 75 & 43   \\
% \hline               
0.002 &  37 & 55 & 105   &  35  & 100 &  48 \\
\hline
\end{tabular}}
\end{center}
\tablefoot{
\tablefoottext{a}{non-rotating BPASS V2.0 single-star tracks \citet{Eldridge2008, Eldridge2016}}
\tablefoottext{b}{rotating ($v_{\rm rot, i}=40\%$ critical) Geneva tracks, \citet{Georgy2015}}
\tablefoottext{c}{rotating ($v_{\rm rot, i} = 150\,$\kms) FRANEC tracks, \citet{Limongi2018}. Values for 
$Z{=}0.006$ are obtained through linear interpolation between $Z{=}0.002, 0.014$.}
\tablefoottext{d}{non-rotating FRANEC tracks \citet{Limongi2018}}
\tablefoottext{e}{non-rotating STARS tracks, \citet{Dray2003}}
\tablefoottext{f}{These tracks imply self-stripping already at initial masses of ${\approx} 15\,M_\odot$. However, following our 
strict definition, \Mtwo~cannot be smaller than \Mone, and hence we set \Mtwo=\Mone.}
}
\end{table}

\section{Estimating \Mtwo: At what masses can single stars undergo self-stripping?}
\label{sec:EstimatingM2}

The parameter \Mtwo~has been extensively discussed in the literature. Up until a decade or so, 
a typical value for this parameter at $Z = Z_\odot$  
used to be cited as $\approx 25\,M_\odot$ \citep[e.g.][]{Crowther2007}. Naive estimations 
of this parameter relied explicitly on calibration to the 
lowest-luminosity apparently-single WR star. However, apparently-single WR stars may still have experienced binary interaction, and there are 
multiple scenarios to support this idea. To name a few: 1) The envelope of the WR progenitor could have been stripped during 
a second Roche lobe overflow phase onto a compact object (the original primary), and the once high-mass X-ray binary is now 
a long period WR + compact object binary in a quiescent state that escaped detection 2) The progenitor of the 
WR star may have been stripped by a compact object or a lower mass object 
during common envelope evolution \citep[][]{Paczynski1976, Schootemeijer2018}, where the lower-mass companion 
may have escaped detection; 3) After being stripped, the WR star was kicked away 
through three-body interactions in triple systems, becoming a single WR star with a history of binary interaction; 
4) Some WR stars could be evolved merger products. The merging process may result in 
enhanced mass-loss due to super-Eddington winds and eruptions  \citep[e.g.,][]{Owocki2017} and 
enhanced rotation \citep[e.g.][]{DeMink2013}. 
However, models calculated by \citet{Schneider2019} 
suggest the presence of strong magnetic fields in massive mergers, which may suppress both these phenomena.
Whether a merger is more likely to become a WR star or not is therefore not conclusive.

Hence, calibrating evolution models to apparently-single WR stars can be very risky.
% \footnote{One idea worth pursuing is to identify 
% a WR binary with the lowest-luminosity WR star in which the companions avoided interactions and evolved as stars in 
% isolation. However, this relies on our ability to establish the evolution history of the binary with certainty, which is only rarely feasible 
% \citep[e.g.,][]{Shenar2019}.} 
Incidentally, we note that in all three galaxies, the lowest-luminosity 
WR stars are apparently single stars (Fig.\,\ref{fig:WRpops}) . If we were to calibrate \Mtwo~according to them, we would naturally obtain \mbox{\Mone = \Mtwo}, and 
the binary channel would be deemed irrelevant. However, there is no reason to suspect that 
the minimum mass necessary to appear 
as a WR star (\Mone) is identical to the minimum mass necessary for a star to strip itself (\Mtwo) in all three galaxies.
Thus, \Mtwo~cannot be derived from the empirical distribution of WR stars, and one needs to rely 
on the predictive power of evolution models.

The ability of a star to strip itself depends on two main factors: the mass-loss history of the star prior to the cWR phase (including eruptions), and the size 
of its convective core (mixing, rotation), including the interplay between the two.
Needless to say, both these domains are hampered with uncertainties. Therefore, instead of trying to determine specific values 
for \Mtwo, we examine predictions by various evolution codes. 
In Table\,\ref{tab:M1M2}, we summarize \Mtwo~values obtained 
from the BPASS V2.0 \citep{Eldridge2008, Eldridge2016}, FRANEC \citep{Limongi2018}, 
Geneva \citep[][Eggenberger et al.\ in prep.]{Ekstrom2012, Georgy2012, Georgy2015}, and
STARS \citep{Dray2003} stellar evolution codes, using various assumptions on rotation.

Values of \Mtwo~are not always  stated by the authors. In these cases, we identify the lowest initial 
mass at which self-stripping occurs. We then interpolate between this value and the next-lowest initial mass in the corresponding 
grid to estimate \Mtwo. For example, as the $Z=0.002$ BPASS single-star track for $M_{\rm i} = 50\,M_\odot$ does 
not undergo self-stripping and $M_{\rm i} = 60\,M_\odot$ track does, we estimate \Mtwo$=55\,M_\odot$. We also 
note that the cWR phase is only considered to be reached if \mbox{$T_* > 30\,$kk} for the post main-sequence star. 
As can be seen from Fig.\,\ref{fig:WRpops}, 
no WR stars with $\log T_* \lesssim 30\,$kK are known. This is different than the convention used by
\citet{Georgy2012}, for example, who defined the WR phase at $T_* \ge 10\,$kK. Hence, the \Mtwo~values 
given here may slightly deviate from values reported by the respective authors (e.g., compare Table\,\ref{tab:M1M2} with 
\citealt{Georgy2015}).

The differences between the codes are abundant, and it is beyond the scope of our paper to discuss them in detail. 
Rather, Table\,\ref{tab:M1M2} serves to give an impression of the uncertainties that dominate the parameter \Mtwo, which become 
most extreme at low $Z$. For example, rotating FRANEC tracks at $Z=0.002$ get to the cWR phase\footnote{According to \citet{Limongi2018}, 
at metallicities
of roughly $Z=0.002$ and lower, dust-driven winds during the 
red supergiant  phase dominate the mass removal, making \Mtwo~independent of $Z$.} already 
at $35\,M_\odot$, 
while rotating Geneva tracks barely reach the cWR phase even for the highest masses considered (120$M_\odot$). 
Instead of considering which of these predictions is closer to reality, we instead illustrate their consequence regarding the 
relative contribution of the binary channel.

\section{The ``window of opportunity'' for the binary channel, [\Mone, \Mtwo]}
\label{sec:bincont}

\begin{figure*}[]
\minipage{0.49\textwidth}
  \includegraphics[width=1\linewidth]{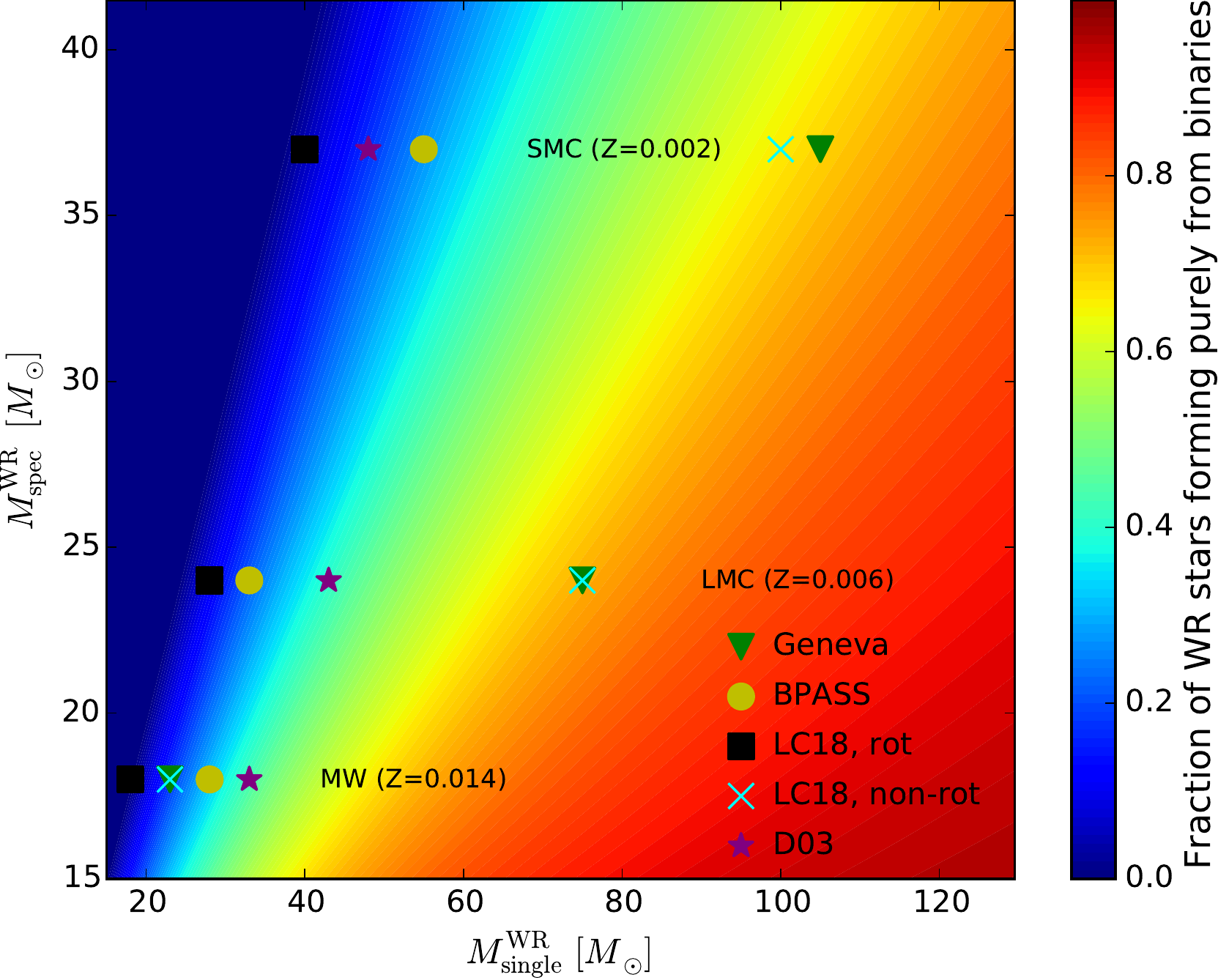}
%   \caption{A really Awesome Image}
%   \label{fig:awesome_image1}
\endminipage\hfill
\minipage{0.49\textwidth}
  \includegraphics[width=\linewidth]{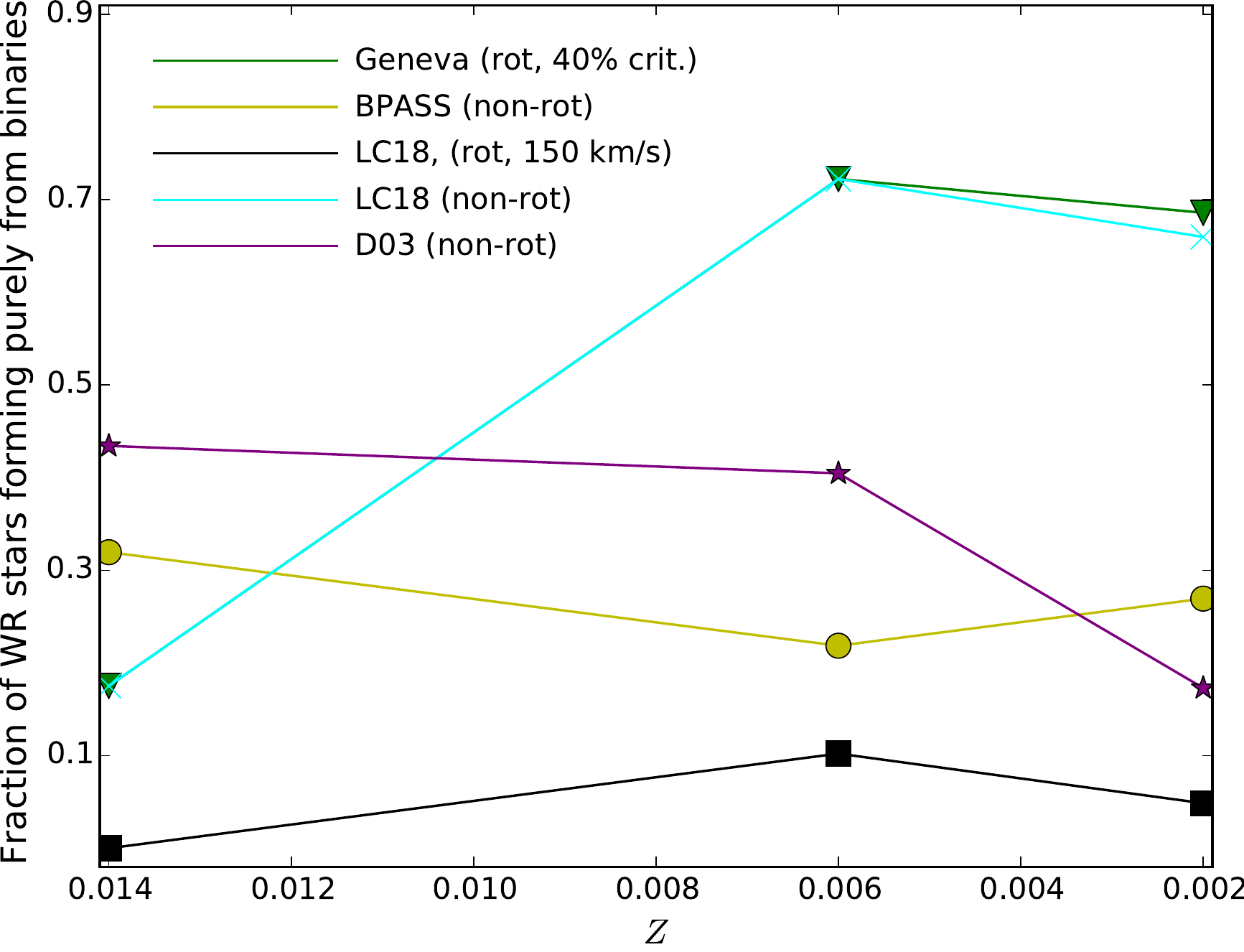}
%   \caption{A really Awesome Image}
%   \label{fig:awesome_image2}
\endminipage
\caption{\emph{Left:} A color map depicting the relative contribution of the binary channel (Eq.\,\ref{eq:bsWR}) on the \Mone-\Mtwo~plane. 
Marked are the relative contributions obtained for the \Mone~values derived here for $Z=0.014, 0.006, $ and 0.002 ($18, 24, 37\,M_\odot$) 
and the \Mtwo~values predicted by the BPASS, Geneva, FRANEC, and STARS evolution codes (cf.\ Table\,\ref{tab:M1M2}). 
\emph{Right:} A translation of the left panel to the $Z$-axis, showing the relative contribution of the binary channel (Eq.\,\ref{eq:bsWR}) as a function of $Z$
according to the various evolution models.
% as predicted by 
% assuming the values of \Mone~derived here and \Mtwo~derived from various evolution codes.
}
\label{fig:fracs}
\end{figure*}

Binary stripping can primarily alter the number of WR stars in a given population for stars in the mass range \Mone${<}M_{\rm i}{<}$\Mtwo:
Stars initially less massive than \Mone~would not appear as WR stars after stripping.
In contrast, stars more massive than \Mtwo~would undergo self-stripping and rapidly enter the cWR phase as single stars, regardless of binary 
interactions. While the time-scales of the stripping could be slightly altered by the presence of a companion, the difference is negligible compared to the 
lifetime of the WR phenomenon, as is evident from the lack of red and yellow supergiants that correspond to $M_{\rm i} \gtrsim 30\,M_\odot$ 
\citep[e.g.,][]{Davies2018}.
The interval [\Mone($Z$), \Mtwo($Z$)] can therefore be considered as a ``window of opportunity'' for binary interactions 
to affect the WR population. 

A few conclusions can be drawn immediately, just from the existence of this interval. For example, 
if for some reason \Mone = \Mtwo, that is to say, stars that appear as WR stars after stripping are stars that could strip themselves, 
then the additional contribution of the binary channel to the number of WR stars would vanish. If, in contrast, \Mone $\ll$ \Mtwo, then 
the additional contribution would approach 100\%, since virtually no WR stars could form as single stars except for perhaps 
the most massive stars.

We can quantify the additional contribution of the binary channel as follows. Using the terminology introduced by \citet{Shenar2019}, let us 
use the designation b-WR star for a WR star that could only form through binary interactions, that is, a WR star 
with an initial mass in the interval [\Mone, \Mtwo]. Then the fraction of b-WR stars of the whole WR population would be:

\begin{equation}
\resizebox{0.5\textwidth}{!} 
{$
% a + b - \frac{a}{b}
\frac{N_\text{b-WR}}{N_\text{cWR}}(Z) = 
\frac{\,\int\limits_{M_{\rm spec}^{\rm WR}(Z)}^{M_{\rm single}^{WR}(Z)} f_\text{strip}\,m^{-2.35}\,t_\text{WR}(m)\,{\rm d}m}
 {\int\limits_{M_{\rm spec}^{\rm WR}(Z)}^{M_{\rm single}^{WR}(Z)}f_\text{strip}\,m^{-2.35}\,t_\text{WR}(m)\,{\rm d}m + 
 \int\limits_{M_{\rm single}^{WR}(Z)}^{M_{\rm i, max}}  m^{-2.35}\,t_\text{WR}(m)\,{\rm d}m}.
%  \int_{M_\text{i,w-WR}}^\infty} m^{-2.35}\,T_\text{cWR}(m)\,{\rm d}m} \approx 0.7, 
$}
\label{eq:bsWR}
\end{equation}
Here, $N_\text{b-WR}/N_\text{cWR}$ is the fraction of WR stars that formed exclusively via the binary channel, $f_{\rm strip}$ is the 
fraction of stars successfully stripped by their companion, $t_{\rm WR}$ is the lifetime of the cWR phase, 
and  $M_{\rm i, max}$ is the upper mass limit of stars. The
Salpeter initial mass function is assumed \citep{Salpeter1955}. We adopt $f_{\rm strip} = 0.33$ \citep{Sana2012}, 
and fix $t_{\rm WR}$  to the He-burning lifetime \citep[][eq.\ 79]{Hurley2002}. We assume $M_{\rm i, max} = 300\,M_\odot$ 
\citep[e.g.][]{Crowther2010}, but note that the differences are indiscernible for $M_{\rm i, max} \gtrsim 200\,M_\odot$. 

In the left panel of Fig.\,\ref{fig:fracs}, we show the values of Eq.\,(\ref{eq:bsWR}) on the \Mone-\Mtwo~plane. 
We see that for \Mtwo $\gg$ \Mone, 
the additional contribution of binaries to the WR population approaches unity, while for \Mtwo $\approx$ \Mone, 
it vanishes. 
On the same plot, we mark the coordinates that correspond to the different 
predictions using the different codes, given in Table\,\ref{tab:M1M2}.
In the right panel of Fig.\,\ref{fig:fracs}, we plot the respective fractions as a function of $Z$. 
No coherent trend arises from this plot, which is not surprising considering the vast differences in the predictions of the different 
codes. Moreover, none of the curves is monotonically increasing with decreasing $Z$. An exception to this is obtained when 
one assumes an upper mass limit of $150\,M_\odot$, in which case $N_\text{b-WR}/N_\text{cWR}$ in the Geneva 
models reaches 0.18, 0.77, and 0.80  in the MW, LMC and SMC, respectively (i.e., very weakly monotonic).

The FRANEC rotating models are capable of stripping single stars at rather low masses, which is why the ``window of opportunity'' for 
binaries to take action is relatively small. At $Z = Z_\odot$, rotating FRANEC tracks can actually strip at initial masses lower 
than \Mone, which leaves no room for binaries to affect the number of WR stars in the MW. If true, this means that the Galaxy 
contains self-stripped stars that do not appear as WR stars, for which clear evidence still lacks. At $Z=Z_{\rm LMC}$, \Mtwo~is 
slightly larger than \Mone, leaving some room for binaries to contribute at a 10\% level. Going further to $Z=Z_{\rm SMC}$, 
the binary contribution remains modestly low, and even slightly decreases. Hence, assuming the FRANEC predictions are representative 
of realistic \Mtwo~values, 
the additional contribution of binary interaction to the number of WR stars is virtually $Z$-independent, and is generally very low. 
The non-rotating FRANEC models are generally less successful in stripping stars, and result in a non-trivial 
dependence of the relative contribution of the binary channel on $Z$, with b-WR fractions ranging between 20-70\%.

The Geneva rotating models  behave  similarly to the non-rotating FRANEC tracks, predicting at first an increase of the 
additional contribution of the binary channel from MW to LMC, and then a decrease. We note, however, 
that the initial rotation of 40\% critical adopted by the Geneva models 
does not seem to agree with empirical measurements of rotational velocities of O-type stars \citep[e.g.][]{Ramirez2013, Ramirez2015}. 
The BPASS tracks imply a somewhat intermediate behavior, with the relative contribution of the binary channel dropping between 
the MW and the LMC, and then increasing again at the SMC. Meanwhile, the STARS code 
implies a \emph{decrease} of the additional  contribution with decreasing $Z$ - the opposite of the usual claim. 

Despite these contradictory results, the conclusion  is simple: the impact of binary 
interactions in forming 
WR stars depends on $Z$ in a highly non-trivial manner. The additional contribution of the binary channel may 
increase or decrease with $Z$, depending on the metallicity regime. Hence, one should not expect 
a-priori that, at low $Z$, binaries are more important in forming WR stars, or that the 
WR binary fraction must increase.

\section{Disclaimers}
\label{sec:Disc}

\subsection{The apparent lack of a distinct mass interval}
\label{subsec:lackobs}

In our work, we motivate the existence of an initial-mass interval [\Mone,\Mtwo] in which  
WR stars can only form through binary interactions.  
The existence of this interval is the result of 
two seemingly straightforward facts, namely, the existence of \Mone~(motivated in Sect.\,\ref{subsec:Moneex}) and \Mtwo~(a consensus
in the literature). Yet the distributions of the apparently-single and binary WR stars in the SMC and LMC
do not suggest the presence of such an interval. In fact, the SMC shows an opposite picture, where the low-luminosity 
WR stars appear to be single. From this, we conclude that at least one of the following 
should hold:

\begin{enumerate}
 \item \emph{Binary evolution in disguise:} The low-luminosity apparently-single WR stars may be products 
 of binary evolution \citep[e.g.,][]{Schootemeijer2018}, 
 as we thoroughly discussed in Sect.\,\ref{sec:EstimatingM2}. The apparent lack of a pure binary interval 
 \mbox{[\Mone, \Mtwo]} may be due to falsely assuming the apparently single stars are not the products of binary interaction.
%  interpreting apparently-single WR stars 
%  as stars without a history of binary interaction.
 
 \item  \emph{Undetected binaries:} Due to observational biases,
 a population of WR binaries with bright mass-gainer companions may have avoided detection. 
 
 \item \emph{\Mone${\approx}$\Mtwo:} It is possible that, in the SMC and LMC at least,
 stars that appear as WR stars after stripping are stars that undergo self-stripping. As this is
 not typically the prediction of evolution codes (cf.\ Table\,\ref{tab:M1M2}), this could imply
 (I.) underestimated mass-loss rates (including eruptions); (II.) underestimated mixing (due to rotation 
 or otherwise).
 
 \item \emph{Overestimated binary stripping:} The initial conditions and detailed evolution of WR progenitors, which 
 tend to be at the upper-mass range, are not well established. It is possible that the binary-stripping efficiency 
 adopted here ($f_{\rm strip} = 0.33$) is overestimated (e.g., because of decreased likelihood of interaction, or increased 
 likelihood of merging, at the upper-mass range). While this alone does not help explain the existence of low-luminosity WR stars, 
 it does help explain the apparent lack of low-luminosity WR binaries.
\end{enumerate}

It is beyond the scope of the paper to investigate which of these scenarios is more likely to hold. More surveys 
of the WR stars will be needed to confirm or rule out the presence of hidden companions or additional WR binaries in the 
Magellanic Clouds. Moreover, a realistic treatment 
of rotation, binary interactions, and triple-body interactions should be implemented in evolution and  
models 
to examine whether these mechanisms can produce apparently-single low luminosity WR stars.

Regardless of which scenario holds: even if the interval 
[\Mone, \Mtwo] does not exist or is very small, it would still be consistent with binary interactions 
not becoming increasingly important in forming WR stars at low metallicities.

\subsection{Uncertainties rooted in rotation}
\label{subsec:rot}

Rapid initial rotation tends to increase the size of the convective core 
and potentially boost the stellar mass-loss \citep[e.g.,][]{Maeder2000}. 
% {\bf As an extreme case, } near-critically rotating 
% stars, especially at low metallicity, are thought to evolve homogeneously, potentially reducing both the initial-mass 
% limit needed to appear as a WR star (\Mone) and to becomes a WR star (\Mtwo) through enhanced mass-loss, increased $L/M$, and 
% increased temperatures \citep{Szecsi2015, Kubatova2019}. For example, chemical homogeneous evolution was
% invoked to explain the existence of apparently-single WR stars in the SMC \citep{Hainich2015, Ramachandran2019}.
% However, \citet{Schootemeijer2017} demonstrated that their properties are not consistent with homogeneously-evolving stars.
Therefore,
enhanced rotation would generally tend to decrease the values of \Mone~and \Mtwo~in a non-trivial 
manner. Stars at lower $Z$ are known to exhibit faster rotation on average \citep[e.g.][]{Ramirez2013, Ramachandran2019}. 
It is therefore possible 
that distinct spin distributions at different metallicities  indirectly impact  
the formation processes of WR stars as a function of $Z$.

It is difficult to predict the consequences this would have on the importance of binary interactions in forming 
WR stars. On the one hand, the effects of rotation become more prominent at low $Z$, which should generally 
increase the efficiency of the single-star channel in forming WR stars. 
On the other hand, it is possible that accretion of angular momentum during binary mass transfer  
contributes to the mixing of the mass gainer \citep[e.g.,][]{Eldridge2017}. If so, binary interactions may contribute to the 
formation of WR stars by forming rapidly rotating 
mass accretors (rather than through the stripping of the primaries alone). However, it is questionable whether 
mass-transfer can indeed efficiently mix the star after a chemical gradient has already been established.

All in all, the effects of rotation are extremely difficult to account for consistently given the multitude of 
mixing prescriptions and their interplay with binary interaction and mass-loss. This 
only stresses that the importance of binary interactions in forming WR stars is a
non-trivial and strongly model-dependent variable.

\subsection{Why are binary interactions important}
\label{subsec:binimp}

The fact that the binary channel  is not necessarily increasingly important for forming WR stars at low $Z$ does not mean 
that binary interactions in general are not. After all, it is clear that at low $Z$, mass-loss rates are smaller.
How small they become is still debatable, 
since it is possible that $Z$-independent processes such as dust-driven winds  \citep{VanLoon2000} or
eruptions \citep{Smith2014, Owocki2017} dominate the mass-loss at low $Z$. However, it is clear that, at low $Z$, 
binaries are expected to be an increasingly dominant agent for stripping.
Hence, while we have shown that  binary interactions may have a
limited impact on the formation of WR stars (i.e., stars with a WR spectrum), 
it may still be the main channel with which He stars in general are formed.
% it can have a very significant impact on, e.g., the ionizing radiation supplied by stars 

Moreover, one should not confuse the incidence of binary interaction among WR stars with the
additional contribution of the binary channel 
to the number of WR stars. 
Even if the additional contribution were low, it is still possible that the majority of WR stars 
interacted with a companion \citep[e.g.,][]{Vanbeveren1980}.
% After all, the majority of massive stars will interact with a companion during their lifetime. 
This may slightly affect the WR population:
the presence of a companion may prevent the WR progenitor from becoming a red supergiant, 
and may therefore lead to longer WR lifetimes. 
We note, however, that the blatant lack of red supergiants with $M_{\rm i} \gtrsim 25\,M_\odot$
\citep[e.g.][]{Humphreys1979, Davies2018} implies that WR progenitors spend a very short amount of time in this phase or skip it 
altogether.
% While having limited impact on formation of WR stars, mass-transfer has dire consequences on the mass gainer, 
% as well as on the evolution of the binary system as a whole.
% It is worthwhile noting, however, that the incidence of binary interaction among WR progenitors is not well constrained, since it is not clear 
% how large these progenitors become. The existence of the Humphreys-Davidson limit \citep{Humphreys1979, Davies2018} implies that 
% WR progenitors may avoid expansion to very large radii, thereby severely reducing the likelihood of interaction. 

Lastly, it is important to stress that this work does not ``prove'' that the additional contribution of the binary formation channel to the formation 
of WR stars is negligible. In principle, it only shows that there is no reason a-priori to believe that its influence
on the WR population grows with decreasing $Z$. 
% However, considering our results, as well as the similar propert
% values of \Mtwo~predicted by the various evolution codes generally imply that the binary 
% channel is of secondary importance in creating WR stars, with the exception of the Geneva rotating models at low $Z$. Moreover, the similar 
% HRD positions of apparently-single and binary WR stars \citep{Massey1981, Shenar2016, Shenar2019} do not give 
% clear evidence for the importance of the binary channel. 
% The fact that the lowest luminosity WR stars in all Galaxies appear to be single is extremely interesting. 
% This could mean that we strongly underestimate mass-loss or mixing in stars with $M_{\rm i} \approx 20-40\,M_\odot$. 
% Alternatively, it could mean that all these apparently-single WR stars are, in fact, products of binary evolution 
% (see Sect.\,\ref{sec:EstimatingM2}). Another peculiarity is the apparent scarcity of low-luminosity WR binaries. After all, 
% this is where we can naively expect WR binaries to be most common! This scarcity either implies that we are missing on 
% many WR binaries in these galaxies, or that the efficiency of binary stripping in the mass 
% interval $M_{\rm i} = 20-40\,M_\odot$ is strongly overestimated. More effort should be 
% dedicated into answering these questions.

\section{Summary}
\label{sec:summary}

In this work, we addressed the question of whether the WR binary fraction is truly expected to increase with decreasing $Z$, 
as is commonly claimed in the literature. We argued that stars with initial masses below a certain 
mass threshold \Mone, which is a monotonically increasing function of $Z$, would not appear as WR stars spectroscopically. 
Using empirical HRD positions of WR stars in the SMC, LMC, and MW, we constrained minimum initial masses 
of \Mone $= 18, 24, $ and $37\,M_\odot$ for $Z = 0.014, 0.006,$ and $0.002$ (MW, LMC, SMC), respectively. Hence, for example, a 
star with an initial mass of $M_{\rm i} = 30\,M_\odot$ would probably not 
appear as a WR star in the SMC after being stripped by a companion, but would do so 
in the Galaxy.

We continued by compiling the various predictions that exist for the well-known parameter \Mtwo - the minimum initial mass above which 
stars reach the cWR phase through self-stripping. We relied on the FRANEC, BPASS, Geneva, and STARS evolution codes.
The predictions of \Mtwo~vary strongly 
between the codes, especially at low $Z$, and illustrate the large uncertainties involved in its calculation (see Table\,\ref{tab:M1M2}).

We argued that the binary channel can form additional WR stars primarily 
in the initial mass interval [\Mone, \Mtwo]. 
As both \Mone~and \Mtwo~grow with 
decreasing $Z$, the additional contribution of the binary channel depends in a non-trivial manner on $Z$. Using the estimated 
values for \Mone~and \Mtwo for three distinct metallicities, and weighing against the initial mass function and the WR lifetime
(Eq.\,\ref{eq:bsWR}), 
we could produce predictions for the relative contribution of the binary channel. 
% {\bf While the mass interval [\Mone, \Mtwo] is not directly observed, we discussed 
% in Sect.\,\ref{subsec:lackobs} why this may be the case, and why this does not interfere with our conclusions.}

Even though the results are heavily code-dependent and potentially affected by observational biases  (see Sects.\,\ref{subsec:lackobs}
and \ref{subsec:rot}), one result emerges: 
within current uncertainties, no 
model can be claimed to conclusively predict a monotonically increasing behaviour (in fact, 
one of the models implies the opposite trend, see Fig.\,\ref{fig:fracs}). 
In light of the uncertainties, we cannot rule out that binary
interactions become increasingly important in forming WR stars at
low $Z$. Nonetheless, we have demonstrated in this paper that, contrary to
common belief, this is not given a-priori. Thus, one should not 
prematurely expect that 
binary interactions dominate the formation of WR
stars at low metallicity, or that the WR binary 
fraction should increase with decreasing $Z$.

\begin{acknowledgements}
We would like to thank our anonymous referee for their very critical and insightful comments 
- they have greatly contributed to our work.
T.S.\ acknowledges support from the European Research Council (ERC) under the European Union's 
DLV-772225-MULTIPLES Horizon 2020 research and innovation programme. 
A.A.C.S.\ is supported by STFC funding under grant number ST/R000565/1.
T.S.\ would further  like to thank insightful comments and inspiring conversations with A.\ F.\ J.\ Moffat, W.-R.\ Hamann, 
M.\ Limongi, R.\ Hainich, H.\ Todt, L.\ M.\ Oskinova, \& N.\ Langer.

\end{acknowledgements}

\bibliography{literature}

\begin{appendix}
% % 
\section{Deriving \Mone~from \Lone}
\label{sec:M1ini}

To estimate the initial mass \Mone~that corresponds to the luminosity of the He star \Lone,
we use the \texttt{binary} module of Modules for Experiments in Stellar Astrophysics code 
(\texttt{MESA}, version 10398, \citealt{Paxton2011,Paxton2013,Paxton2015,Paxton2018}) 
to evolve stellar models with $33$ different zero-age main sequence (ZAMS) primary masses between 
$M_\mathrm{ZAMS}=10\,\mathrm{M}_\odot$ and $M_\mathrm{ZAMS}=107\,\mathrm{M}_\odot$ with each model having a 
ZAMS secondary mass of half the primary mass. For the initial orbital period we use $12$ values between 
$P_\mathrm{i}=3\, \mathrm{d}$ and $P_\mathrm{i}=10\,000\, \mathrm{d}$ with logarithmic spacing. 
Two metallicity values are used, appropriate for the SMC and LMC. In total, $792$ binary evolution tracks are generated.

\begin{figure}[h]
\centering
\includegraphics[width=0.5\textwidth]{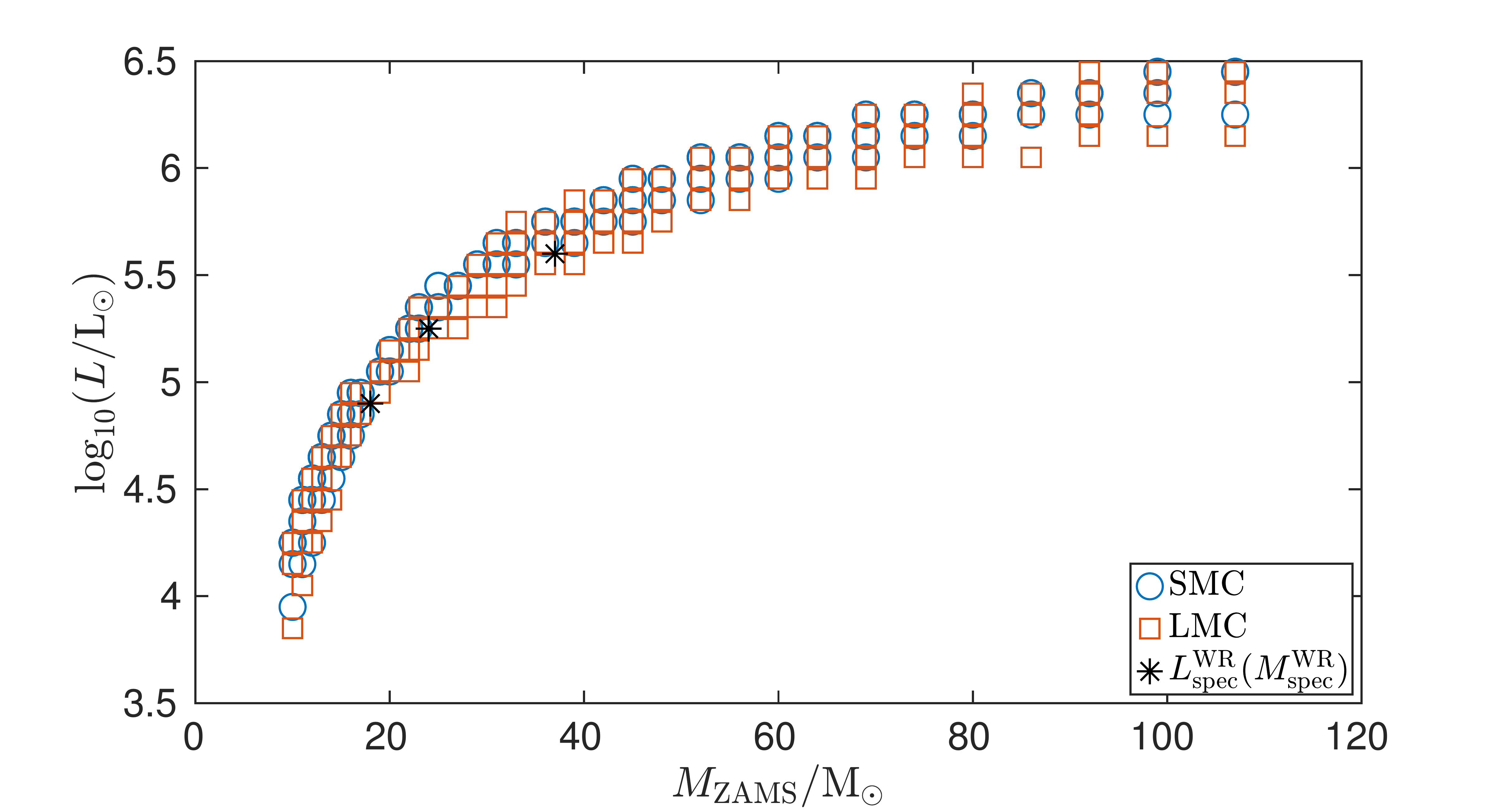} \\ 
\caption{Luminosity bin in which each stellar model spent the longest time in its hot post-MS stage for SMC models (blue circles) and LMC models (orange squares). Our choice for $M_\mathrm{spec}^\mathrm{WR}$ and $L_\mathrm{spec}^\mathrm{WR}$ is marked by black asterisks.}
\label{fig:ML1}
\end{figure}

Convective mixing employs a mixing-length parameter of $\alpha_\mathrm{MLT}=1.5$. Semiconvective mixing follows \cite{Langer1983} 
with an efficiency parameter of $\alpha_\mathrm{sc}=1$. We employ step overshooting above convective cores with $\alpha_\mathrm{ov}=0.335$.
All models are rotating with an initial rotation velocity of $V_\mathrm{i} = 100\,\mathrm{km}\,\mathrm{s}^{-1}$, with the 
implementation of rotation in MESA described by  \cite{Paxton2013}, and the calibration of the mixing efficiency of \cite{Heger2000I}. 
Transport of angular momentum because of magnetic torques is according to \cite{Fuller2019}.

Mass loss by stellar winds follows \cite{Vink2001} when $T_\mathrm{*}>10\,$kK and $X_\mathrm{H}>0.4$.
For $T_\mathrm{*}<10$\,kK the mass-loss prescription of \cite{deJager1988} 
is employed. When $X_\mathrm{s} < 0.4$ and the luminosity is
below $L_\mathrm{spec}^\mathrm{WR}\left(Z\right)$ we use the theoretical mass-loss rate of \cite{Vink2017}. 
For higher luminosities and when $X_\mathrm{H} < 0.1$ we follow either \cite{Hainich2014} or \cite{Tramper2016}, 
depending on the surface helium mass fraction (see \citealt{Yoon2017} and \citealt{Woosley2019}). 
Finally, for hot and luminous phases with $0.1 < X_\mathrm{H} < 0.4$, we follow \cite{Nugis2000}.
Mass transfer by Roche lobe overflow follows the prescription of \cite{Kolb1990} with a mass transfer efficiency of zero, 
with the sole purpose of the mass transfer to strip the primary.

\begin{figure}[h]
\centering
\includegraphics[width=0.5\textwidth]{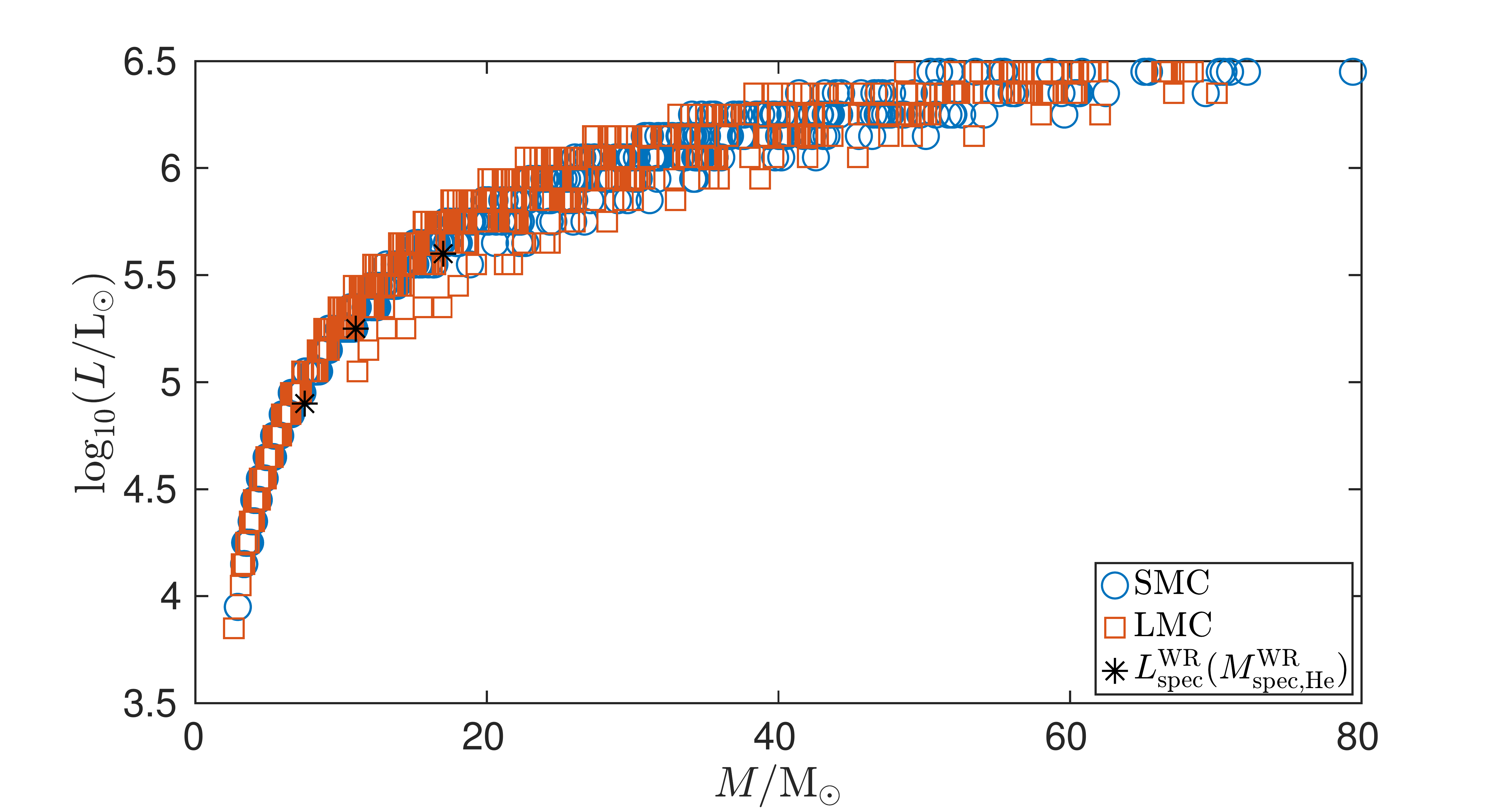} \\ 
\caption{Same as Fig. \ref{fig:ML1}, but with the x-axis showing the stellar mass at the beginning and at the end of the evolutionary phase in which the luminosity is in the longest-duration luminosity bin.}
\label{fig:ML2}
\end{figure}

To get a mass-luminosity relation we find for each model the luminosity bin in which it spends the longest time, 
limiting to hot ($\log T_\mathrm{*} {>} 4.6$[K]) post-MS (central hydrogen mass fraction below $0.01$) 
phases, with each luminosity bin spanning $\Delta \log L{=}0.1\,[L_\odot]$. We find that the duration
spent in the derived luminosity bin is between $40\%$ and $100\%$ of the hot post-MS evolutionary phase we defined, and 
therefore yields a good representation of the mass-luminosity relation. The results are shown in Fig. \ref{fig:ML1}, where 
we also show the $L_\mathrm{spec}^\mathrm{WR}\left(M_\mathrm{spec}^\mathrm{WR}\right)$ values used in the paper.

The method described above directly relates the luminosity of He stars to their ZAMS mass. 
This approach was chosen for simplicity. To conclude, we verify that 
this is consistent with the mass-luminosity relations of \cite{Graefener2011}. 
In Fig.\ \ref{fig:ML2} we plot the mass-luminosity relation obtained from the stellar masses of 
our models when they are in their hot post-MS stage as defined above.

\end{appendix}

\end{document}

%% file: bibdefinitions.tex
\def\apjl{ApJ}%
\def\apjs{ApJS}%
\def\ao{Appl.~Opt.}%
\def\aap{A\&A}%